\documentclass{vldb-clean}

\usepackage{times}

\usepackage{microtype}
\usepackage{graphicx}
\usepackage{booktabs} %

\usepackage{pifont}
\usepackage{hyperref}

\usepackage{paralist}
\usepackage{prp-macros}

\usepackage{rotating}
\usepackage{morefloats}
\usepackage{amsfonts}
\usepackage{yfonts}
\usepackage{upgreek}
\usepackage{tabularx}
\usepackage{xfrac}
\usepackage{cleveref}
\usepackage{balance}
\usepackage{inconsolata}

\usepackage[linesnumbered,vlined]{algorithm2e}

\usepackage{tipa}

\Crefname{equation}{Eq.}{Eqs.}
\crefformat{section}{\S#2#1#3}

\makeatletter
\newcommand{\removelatexerror}{\let\@latex@error\@gobble}
\makeatother

\newcommand{\gpu}{GPU\xspace}
\newcommand{\gpus}{GPUs\xspace}

\newcommand{\gpuvar}{\text{\emph{\textg}}}

\newcommand{\cpu}{\text{CPU}\xspace}

\newcommand{\sys}{\textsc{Crossbow}\xspace}

\newcommand{\sync}{\SMA}

\newcommand{\SMA}{\textsf{\small{}SMA}\xspace}
\newcommand{\easgd}{\textsf{\small{}EA-SGD}\xspace}
\newcommand{\Sync}{\textbf{\large\textsf{SMA}}\xspace}
\newcommand{\tta}{$\mathsf{TTA}$\xspace}

\def\Snospace~{\S{}}

\renewcommand*\algorithmautorefname{Alg.}

\renewcommand\AlCapFnt{\sf\bfseries}

\newcommand{\captext}[1]{\normalfont{}(#1)}

\graphicspath{{.}{../../../scripts/msr/plots/checkpoints/2018-06-21/}{../../../scripts/msr/plots/checkpoints/2018-07-10/}}

\SetCommentSty{mycommfont}

\vldbTitle{}
\vldbAuthors{}
\vldbDOI{} %
\vldbVolume{X}
\vldbNumber{Y}
\vldbYear{2019}

\begin{document}

\title{C\textbf{\Large\textsf{ROSSBOW}}: Scaling Deep Learning with Small Batch
  Sizes\\ on Multi-GPU Servers}

\numberofauthors{3}

\author{
\alignauthor
Alexandros Koliousis\\
\affaddr{Imperial College London}
\and
\alignauthor
Pijika Watcharapichat\\
\affaddr{Microsoft Research}
\and
\alignauthor
Matthias Weidlich\\
\affaddr{Humboldt-Universit\"at zu Berlin}
\and
\alignauthor
Luo Mai\\
\affaddr{Imperial College London}
\and
\alignauthor
Paolo Costa\\
\affaddr{Microsoft Research}
\and
\alignauthor
Peter Pietzuch\\
\affaddr{Imperial College London}
}

\maketitle

\begin{abstract}
  Deep learning models are trained on servers with many \gpus, and training
  must scale with the number of GPUs. Systems such as TensorFlow and Caffe2
  train models with parallel synchronous stochastic gradient descent: they
  process a batch of training data at a time, partitioned across GPUs, and
  average the resulting partial gradients to obtain an updated global model. To
  fully utilise all GPUs, systems must increase the batch size, which hinders
  statistical efficiency. Users tune hyper-parameters such as the learning rate
  to compensate for this, which is complex and model-specific.

  We describe \sys, a new single-server multi-GPU system for training deep
  learning models that enables users to freely choose their preferred batch
  size---however small---while scaling to multiple GPUs. \sys uses many
  parallel model replicas and avoids reduced statistical efficiency through a
  new synchronous training method. We introduce \sync, a synchronous variant of
  model averaging in which replicas \emph{independently} explore the solution
  space with gradient descent, but adjust their search \emph{synchronously}
  based on the trajectory of a globally-consistent average model. \sys achieves
  high hardware efficiency with small batch sizes by potentially training
  multiple model replicas per GPU, automatically tuning the number of replicas
  to maximise throughput. Our experiments show that \sys improves the training
  time of
  deep learning 
  models on an 8-GPU server by 1.3--4$\times$ compared to
  TensorFlow.
\end{abstract}

\section{Introduction}
\label{sec:introduction}

\begin{quote}
  ``If batch size could be made arbitrarily large [...], then training is
  amenable to standard weak scaling approaches.  However, if the training [...]
  is restricted to small batch sizes, then we will need to find other
  algorithmic and architectural approaches to their acceleration.''\\
  \hspace*{\fill}-- J.\ Dean~et al.~\cite{Dean2018}, March 2018
\end{quote}

\noindent
Deep learning has revolutionised many application fields, including computer
vision~\cite{Krizhevsky2012,HeZRS15}, speech
recognition~\cite{Hinton2012,Xiong2016} and natural language
processing~\cite{Johnson2017}. The training of deep learning models is
expensive: it takes roughly half a month to reproduce the state-of-the-art
accuracy for the ImageNet challenge on a single NVIDIA Titan X
GPU~\cite{GoyalDGNWKTJH17}. To reduce training time, systems exploit data
parallelism across many GPUs to speed up training~\cite{Li2014, Ho2013,
  Dean2012}. Consequently multi-GPU servers have become widely available: a new
10-GPU server with NVIDIA Pascal GPUs costs less than \$40,000~\cite{10gpus},
and public cloud providers offer GPU server instances with up to
16~GPUs~\cite{amazon_aws_gpus}.

Users expect training time to go down with the number of GPUs in a
server. Scaling the training process is challenging though because it requires
a system to fully utilise the parallelism of all GPUs without introducing
bottlenecks. Existing systems, including TensorFlow~\cite{Abadi2016osdi},
MXNet~\cite{DBLP:journals/corr/ChenLLLWWXXZZ15},
CNTK~\cite{DBLP:conf/kdd/SeideA16}, and
Caffe2~\cite{DBLP:conf/mm/JiaSDKLGGD14}, use parallel \emph{synchronous
  stochastic gradient descent}~(S-SGD)~\cite{DBLP:journals/corr/Krizhevsky14}
for training: input batches of training data are partitioned across GPUs. Each
GPU then updates its local model replica before a synchronisation operation
calculates a new global model for the training of the next input batch.

To utilise many GPUs effectively, S-SGD must therefore use a large batch
size. The batch size typically grows linearly with (1)~the number of GPUs, and
(2)~the performance of each GPU. In practice, batch sizes of 64,000 are now not
uncommon~\cite{tencent2018}. With large batch sizes though, \emph{statistical
  efficiency}~\cite{Zhang2014b} of the training process
reduces~\cite{Keskar2016, Masters2018}. As the per-GPU model replicas synchronise less
frequently in relation to the processed training data, the converge rate
decreases, which in turn increases the time-to-accuracy until the trained model
reaches a target accuracy. Users try to compensate for this reduction in
statistical efficiency by increasing the learning rate~\cite{GoyalDGNWKTJH17},
or adjusting the batch size adaptively~\cite{Smith2017}. These techniques,
however, require model-specific tuning and do not fundamentally solve the
problem but eventually fail for very large batch
sizes~\cite{Dean2018, tencent2018, GoyalDGNWKTJH17}. Given these implications of large batches, users
prefer to use small batches when possible~\cite{Masters2018}.
 
The goal of our work is to explore how to design a deep learning system that
effectively trains with small batch sizes, \ie between 2 and
32~\cite{Masters2018}, while still scaling to many
GPUs. The starting point for our design is that, on each GPU, we simply train a
model replica with a small batch size.
This introduces two challenges, which we address in the paper: (i)~how to
synchronise this potentially large number of model replicas without adversely
affecting statistical efficiency; and (ii)~how to ensure that the hardware
resources of each \gpu are fully utilised, thus achieving high hardware
efficiency?

We describe the design and implementation of \sys, a new single-server
multi-GPU deep learning system that decreases time-to-accuracy when increasing
the number of GPUs, irrespective of the batch size.\footnote{The open source
  release of \sys is available at:\newline \url{https://github.com/lsds/Crossbow}.} The
design of \sys makes the following new contributions:

\mypar{(1)~Synchronous model averaging~(\SMA)} \sys uses \SMA, a new
synchronisation approach that synchronises model replicas in a scalable fashion
with a low reduction in statistical efficiency. In \SMA, multiple parallel
\emph{learners} each train their own model replica independently. Learners
access a global average model to coordinate: they adjust their trajectories
with an update proportional to their divergence from the average model. The
average model thus reaches better minima faster than individual learners. All
replicas, including the average model, are updated after each learner processes
a single batch, and all accesses to the average model are strongly consistent.

\mypar{(2)~Auto-tuning the number of learners} With a small batch size, a
single learner may not fully utilise the resources of a GPU. \sys therefore
places multiple concurrent learners on the same GPU. The number of learners per
GPU is tuned automatically. During training, \sys increases the number of
learner until there is no increase in training throughput, \ie the maximum
hardware efficiency has been reached. 
It then uses the number of learners that resulted in peak throughput.

\mypar{(3)~Concurrent task engine} \sys has a task scheduler that dynamically
schedules learners that process the next batch on the first available GPU. The
scheduler issues learning and synchronisation tasks concurrently in order to
prevent the synchronisation performed by \SMA from becoming a bottleneck.
\sys achieves this by breaking the global synchronisation barrier of \SMA into
a hierarchical tree: each learner synchronises using a local copy of the
average model that resides on its GPU; and local models synchronise across
GPUs. The local and global synchronisation operations have optimised
implementations according to their communication scopes (\eg using
all-reduce~\cite{sergeev2018horovod}). They also overlap with the forward and
backwards error propagation of learners.

\tinyskip

In our experimental evaluation, we show that, when training ResNet-50 with
2~model replicas per GPU and a batch size of 16, \sys reaches a given target
accuracy 1.5$\times$ faster than TensorFlow. Training with multiple model
replicas per GPU reduces time-to-accuracy by 1.9$\times$ for
ResNet-32, %
by 4.2$\times$ for VGG-16, and by 2.7$\times$ for LeNet, respectively. \sync
improves statistical efficiency with multiple model replicas by up to
1.6$\times$; with multiple model replicas per GPU, the task engine of \sys
improves hardware efficiency by up to 1.8$\times$.
      
The rest of the paper is organised as follows: \S\ref{sec:background} discusses
the challenges when scaling the training by increasing the batch size;
\S\ref{sec:sma} introduces \sys{}'s synchronous model averaging approach with
independent learners; \S\ref{sec:system} describes the design and
implementation of the \sys task engine; \S\ref{sec:evaluation} presents out
experimental results; \S\ref{sec:relatedwork} surveys related work; and
\S\ref{sec:conclusions} concludes.

\section{Scaling Deep Learning}
\label{sec:background}

Deep learning models, \eg multi-layer convolutional neural
networks~\cite{lecun98}, have been shown to achieve high
accuracy for many image or speech classification problems~\cite{HeZRS15,
  ArikCCDGKLMRSS17}. Since increasing the amount of training data and the
number of model parameters improves their accuracy~\cite{HeZRS15, Dean2012},
deep learning models require training approaches that exploit the parallelism
of modern hardware.

\subsection{Mini-batch gradient descent}
\label{sec:ml-background}

Supervised training of deep learning models uses labelled samples, split into
training and test data. A model gradually ``learns'' to predict the labels of
training data by adjusting its model parameters based on the error. It usually
takes several passes (or \emph{epochs}) over the training data to minimise the
prediction error. The test data is used to measure the model accuracy on
previously unseen data. The most important metric is \emph{test accuracy},
which measures the ability of the model to make predictions when deployed ``in
the wild''.

More formally, let $\mathsf{w}$ be a vector of model
parameters~(\emph{weights}), and $\ell_{\mathsf{x}}(\mathsf{w})$ be a loss
function that, given $\mathsf{w}$, measures the difference between the
predicted label of a sample $(\mathsf{x}, \mathsf{y})$ and the ground
truth~$\mathsf{y}$. The training problem is to find a $\mathsf{w^{*}}$ that
minimises the average loss over all training data. In today's systems, this is
achieved using \emph{stochastic gradient descent}~(SGD)~\cite{Robbins1951,
  Bottou1998, Bottou2018}, an iterative training algorithm that adjusts
$\mathsf{w}$ based on a few samples at a time:
\begin{equation}
  \label{eq:iter}
  \mathsf{w}_{n+1} = \mathsf{w}_{n} - \upgamma_{n} \nabla \ell_{\mathsf{B}_{n}}(\mathsf{w}_{n})
\end{equation}
where $\upgamma_{n}$ is the learning rate in the $n$-th iteration of the
algorithm, $\mathsf{B}_{n}$ is a mini-batch of $b$~training samples, and
$\nabla \ell$ is the gradient of the loss function, averaged over the batch
samples:
\begin{equation}
  \label{eq:grad}
  \nabla \ell_{\mathsf{B}_{n}}(\mathsf{w}_{n}) =
  \frac{1}{b} \sum_{\mathsf{x \in B}_{n}} \nabla \ell_{\mathsf{x}}(\mathsf{w}_{n})
\end{equation}

It is common to augment \Cref{eq:iter} with \emph{momentum}, a technique known
to accelerate the convergence of deep learning
models~\cite{Sutskever2013}. Using momentum, the training process favours
gradients that descent in directions known to improve accuracy from previous
iterations. The iterative training
algorithm with momentum becomes:
\begin{equation}
  \label{eq:grad-momentum}
  \mathsf{w}_{n+1} = \mathsf{w}_{n} - \upgamma_{n} \nabla \ell_{\mathsf{B}_{n}}(\mathsf{w}_{n}) + \upmu(\mathsf{w}_{n} - \mathsf{w}_{n-1})
\end{equation}
\noindent where $\upmu$ is the momentum parameter and
$\mathsf{w}_{n}$$-$$\mathsf{w}_{n-1}$ denotes the algorithm's previous search
direction.

Gradient back-propagation~\cite{Rumelhart1986} is used to compute the model
gradients when weights are spread across multiple layers. This is done in two
steps: (i)~an input batch propagates \emph{forward} through the layers to
compute the predicted label. This is compared with the ground-truth label
associated with each sample in the batch, measuring the error; and (ii)~the
error propagates \emph{backwards} from layer to layer in reverse order. The
error is used to compute the gradient for the weights in each layer, and the
weights can then be updated incrementally by applying \Cref{eq:grad-momentum}.

When training a deep learning model, the goal is to reduce the time to reach a
target level of test accuracy~(\emph{time-to-accuracy}). Two factors affect the
time-to-accuracy: (i)~the number of iterations that a training algorithm such
as SGD requires to find a solution with a given test accuracy
(\emph{statistical efficiency}); and (ii)~the execution time of each iteration
(\emph{hardware efficiency}).

\subsection{Training with GPUs}
\label{sec:gpu-background}

GPU architectures are well suited for increasing the hardware efficiency of the
training process. A GPU is a many-core processor that is designed for high
processing throughput. It features thousands of cores, which are simple
floating-point arithmetic units. Arranged in groups, cores form tens of
streaming multi-processors~(SMs). Multiple threads can execute the same
instruction per cycle on different data, such as a training sample or a weight.

When training a deep learning model on a GPU, a batch of training
data~$\mathsf{B}$
(or, during the
backwards phase, its error) is transformed via a series of matrix or vector
floating-point operations (\eg $\mathsf{B} \times \mathsf{w}$) as it propagates
from one layer to the next.
GPUs can perform more floating-point operations per weight read than a
\cpu, thus achieving more model updates per second~\cite{Jouppi2017}.
 
Programs for the GPU are \emph{kernels}, which can be executed in a blocking or
non-blocking fashion. Complex multi-layer deep learning models may comprise of
hundreds of kernels. Kernels are executed in-order as part of a GPU
\emph{stream}, which is a queue of device work. A GPU can have more than one
stream, which allows kernels to execute \emph{concurrently}. Modern GPUs
support \emph{events}, which are a publish/subscribe mechanism to synchronise
across different streams, without having to stall the entire GPU pipeline.

Copying input data from CPU to GPU memory over the PCIe bus is typically
assisted by a \emph{copy engine} on the GPU, which runs independently from the
GPU's compute engine that schedules kernels. Systems therefore hide the latency
of communication by overlapping communication with computation tasks (\eg using
NVIDIA's Collective Communication Library~(NCCL)~\cite{nccl}).

A server can have multiple GPUs, and GPU-to-GPU data transfers use the PCIe bus
or exploit a fast direct interconnect such as NVIDIA's NVLink
bridge~\cite{nvlink}. The GPUs in a server are interconnected in a topology
with varying bandwidth: \eg in a two-socket multi-GPU server, the GPUs may form
a binary tree in which each GPU pair is connected to a PCI switch, and two
pairs are connected with a PCI host bridge, attached to a CPU socket.

\subsection{Parallel synchronous gradient descent}
\label{sec:background:ssgd}

Current parallel training approaches distribute the gradient computation across
multiple GPUs, but differ in how they synchronise the gradients. The prevailing
training algorithm is parallel
\emph{synchronous~SGD}~(S-SGD)~\cite{DBLP:journals/corr/Krizhevsky14}. It
requires all GPUs to have a consistent view of the $n$-th version of the model
before the $(n$$+$$1)$-th iteration starts: (i)~at each iteration, S-SGD
partitions a batch equally across GPUs; (ii)~each GPU computes a \emph{partial}
gradient from a batch partition and the latest model version; (iii)~GPUs then
coordinate to merge partial gradients into an \emph{aggregate} gradient
(according to~\Cref{eq:grad}); and (iv)~the aggregate gradient is used to
update the models (according to~\Cref{eq:grad-momentum}) before the next
iteration.

\begin{figure}[tb]
  \centering
  \includegraphics[width=0.97\columnwidth]{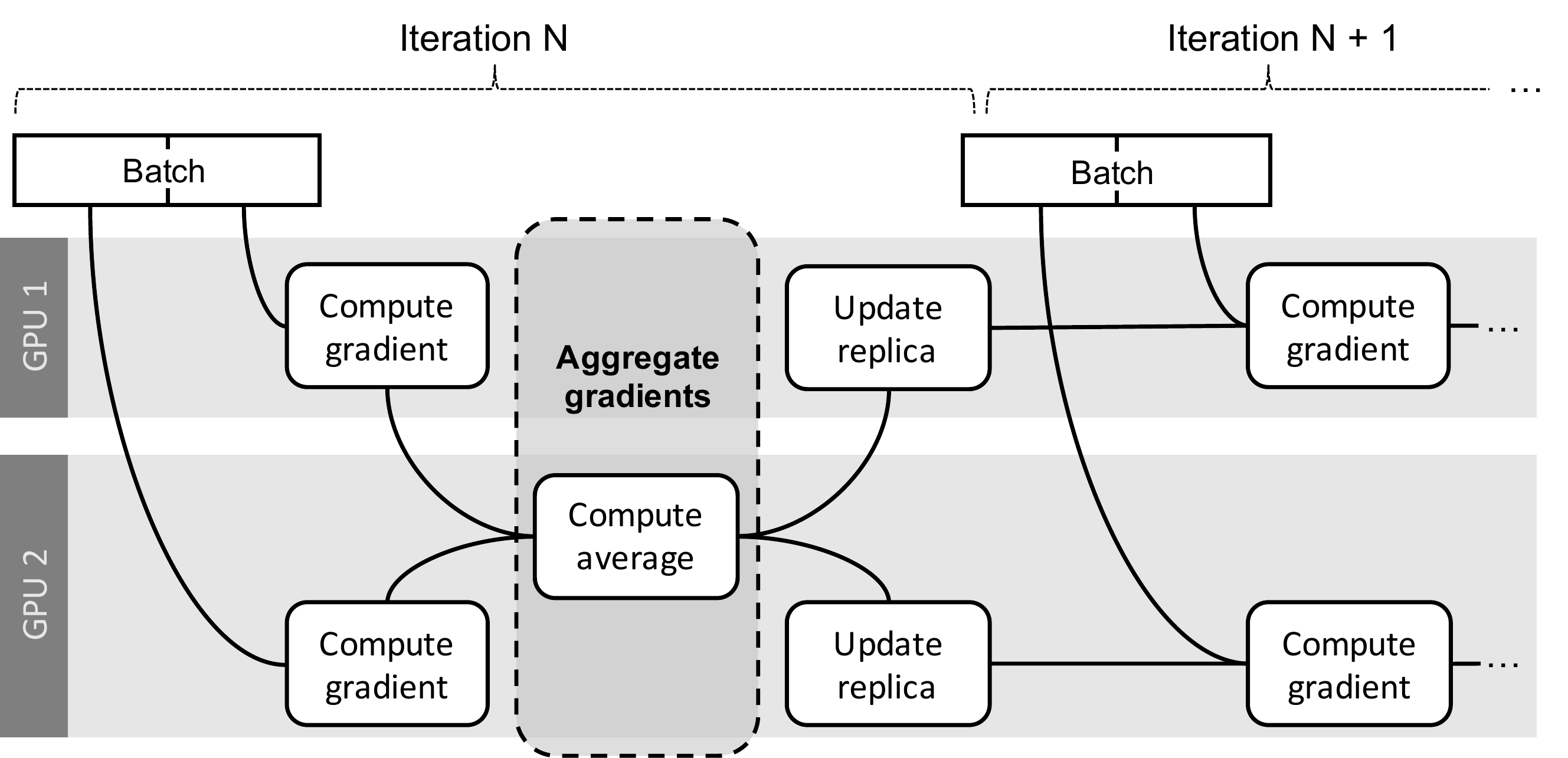}
  \caption{Parallel S-SGD with two GPUs \captext{Each \gpu reads half a batch
      and computes a gradient based on its local model replica. Training does
      not proceed to the next batch until all replicas have been updated with
      the aggregate value of the computed gradients.}}\label{fig:s-sgd}
\end{figure}

\Cref{fig:s-sgd} shows the execution of S-SGD on a two-\gpu server. Each GPU
has a local model replica in its memory, which is used to compute the partial
gradients. The GPUs coordinate so that the same aggregate gradient is applied
to all local replicas ensuring consistency: a \gpu collects partial gradients,
averages them, and then disseminates the result.

To address stragglers during gradient computation or synchronisation,
researchers have proposed an \emph{asynchronous} variant of 
SGD~(A-SGD)~\cite{Chaturapruek2015}. In
A-SGD, a GPU progresses to the next iteration immediately after its partial
gradient was added to the aggregate gradient, and uses the value accumulated
thus far to update its model replica. This leads to \emph{stale} gradients and
hard-to-understand asynchrony, making it difficult to train complex neural
network models such as ResNet effectively. In contrast, S-SGD has better
convergence properties, which is why it has become the de-facto standard for
the training of deep neural networks~\cite{GoyalDGNWKTJH17, tencent2018}. We
therefore also focus on synchronous training.

\subsection{Challenges in scaling training}
\label{sec:problem-background}

\begin{figure}[tb]
  \centering
    \includegraphics[width=.7\columnwidth]{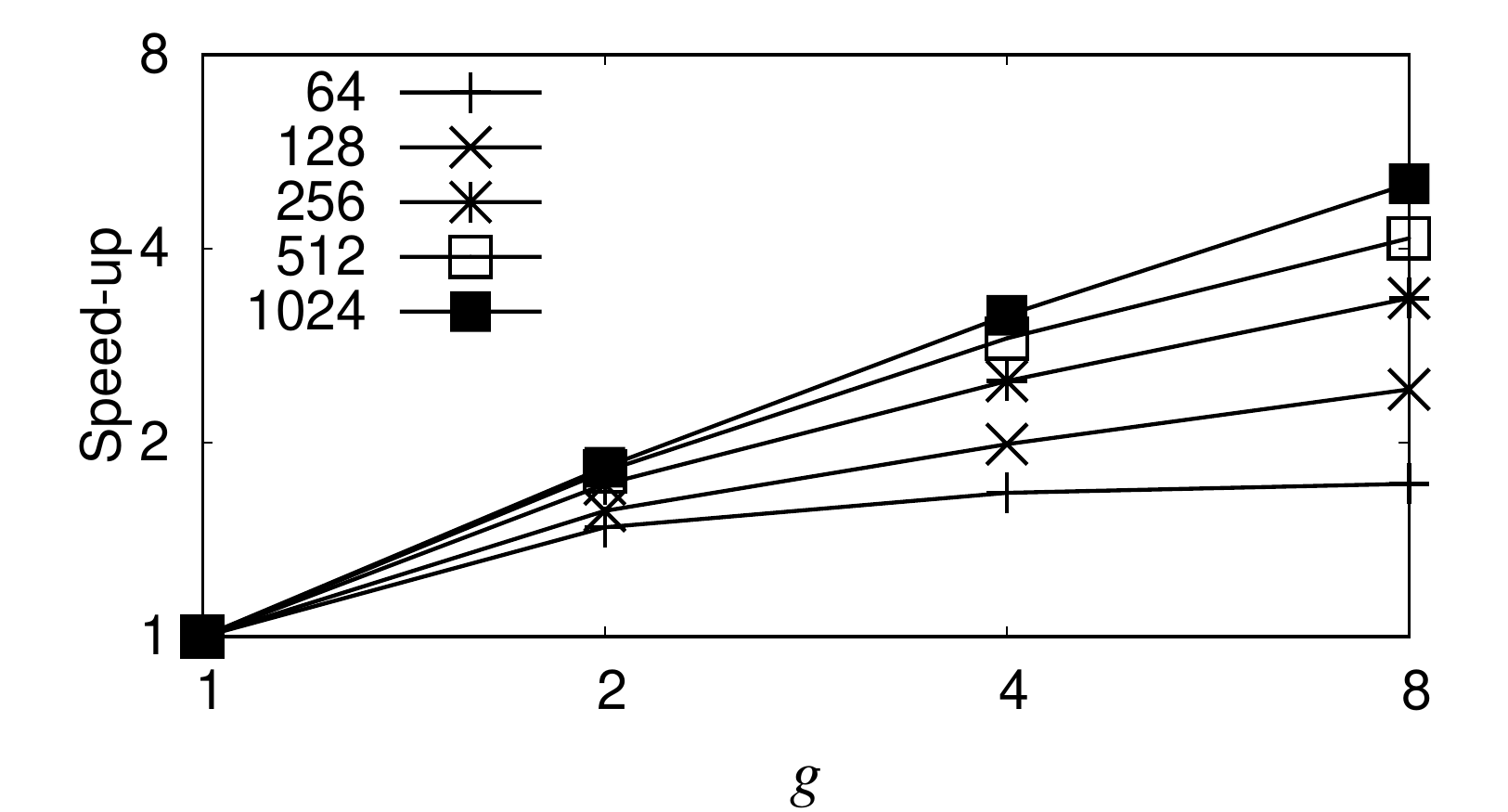}
    \caption{Hardware efficiency \captext{The figure shows the speed-up
        achieved with an increasing number of GPUs, as we vary the batch size
        when training a ResNet-32 model with TensorFlow.}
        }\label{fig:background1}
\end{figure}

\begin{figure}[tb]
  \centering
    \includegraphics[width=.7\columnwidth]{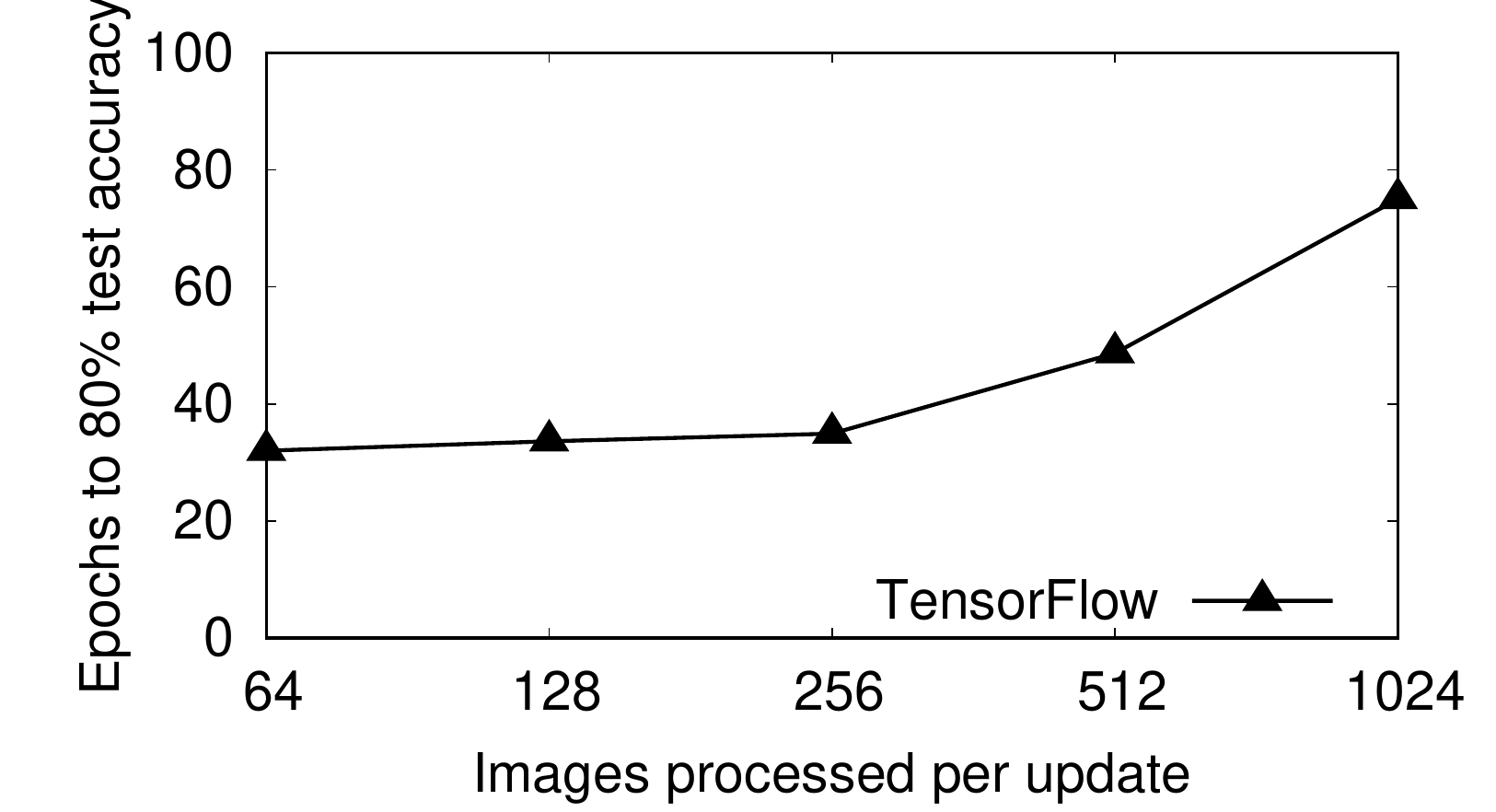}
    \caption{Statistical efficiency \captext{The figure shows the number of
        epochs required to reach a target accuracy of 80\%, as we vary the
        batch size when training a ResNet-32 model with TensorFlow.} 
        }
        \label{fig:background2}
\end{figure}

\begin{quote}
  ``Training with large mini-batches is bad for your health. More importantly,
  it's bad for your test error. Friends don't let friends use mini-batches
  larger than 32.''\\
  \hspace*{\fill}--Y.~LeCun (\texttt{@ylecun}), April 2018
\end{quote}

The batch size is a critical parameter when training with parallel S-SGD: if
the batch is too small, the GPU is not fully utilised as the communication
overhead to move data to and from the GPU dominates. In parallel training, the
(aggregate) batch size must therefore increase \emph{linearly} with the number
of GPUs, resulting in a constant batch size per GPU; otherwise, the overall
throughput scales poorly. This effect can be seen in the plot of hardware
efficiency in~\Cref{fig:background1}. The figure shows the relative throughput
speed-up when training a deep learning model with TensorFlow as we increase the
number of GPUs. If the aggregate batch size remains constant (\eg 64), the
throughput does not increase linearly because the batch size per GPU decreases
(\eg with 8~GPUs the batch per GPU is just 8). In contrast, if we increase the
aggregate batch size (\eg to 512 or 1,024 for 8~GPUs), thus maintaining a
constant batch size per GPU, we observe a linear speed-up.

While the above shows that large batch sizes are ideal to ensure high
\emph{hardware efficiency}, they exhibit poor \emph{statistical
  efficiency}~\cite{Masters2018}, which is expressed as the number of epochs
required to converge to a given accuracy~(ETA). This is shown in the plot of
statistical efficiency in~\Cref{fig:background2}: as the batch size increases,
TensorFlow requires more epochs to converge. The reasons are twofold: (1)~with
large and redundant training datasets (as it is often the case), small batches
ensure faster training because only few batches are sufficient to capture the
dimensionality of the problem space and converge quickly to good
solutions~\cite{LeCun2012, Bottou2018}; (2)~a small batch size leads to
``noisier'' gradient updates, which widen the exploration of the loss
landscape, making it more likely to find better solutions with a higher test
accuracy~\cite{Hoffer2017, Hochreiter1997, Keskar2016, Jastrzebski2017}

This trade-off between hardware and statistical efficiency is particularly
detrimental in parallel training. While increasing the batch size enables a
linear throughput speed-up with the number of GPUs, beyond a certain threshold
(\eg 256 in the example in~\Cref{fig:background2}), the number of epochs
increases \emph{super-linearly}, thus preventing a linear reduction of training
time.

A typical solution to mitigate this issue and compensate for the loss of
statistical efficiency with larger batch sizes is hyper-parameter tuning, \eg
dynamically adjusting the batch size as well as other hyper-parameters, such as
the learning rate and the momentum, during the training progresses. In
particular, it has been observed that, as long as the ratio between the
learning rate and the batch size remains constant, training may be improved by  
varying the batch size~\cite{Jastrzebski2017, Smith2017, GoyalDGNWKTJH17,
  Hoffer2017}. 
This only holds when the learning rate remains relatively small
though~\cite{GoyalDGNWKTJH17}.

While hyper-parameter tuning can achieve quasi-linear scaling of the training
time for large networks such as ResNet-50 on up to
1,024~GPUs~\cite{GoyalDGNWKTJH17, Smith2017, tencent2018}, it requires a
time-consuming model-specific methodology, which is often beyond the reach of
non-experts and cannot be applied easily to new models or hardware
architectures. In some cases, even with hyper-parameter tuning, it is hard to
scale training time on multiple GPUs: a recent study from Google
Brain~\cite{shallue2018measuring} shows that convolutional networks exhibit
only limited scaling with batch sizes larger than 64 and, for recurrent neural
networks, \eg long short-term memory~(LSTM), the threshold seems to be even
lower (16)~\cite{shallue2018measuring}.

A general approach for scaling S-SGD on multiple GPUs therefore remains an open
challenge due to the conflicting impact of large batch sizes on hardware and
statistical efficiency. In the next section, we show how \sys addresses this
problem by leveraging a new synchronisation approach among fully-utilised GPUs,
thus achieving high GPU throughput without sacrificing converge speed.

\newcommand{\HT}{\hspace{-.05em}\raisebox{.3ex}{\normalsize\bf \#}}

\begin{figure}[tb]
  \centering
  \includegraphics[width=0.97\columnwidth]{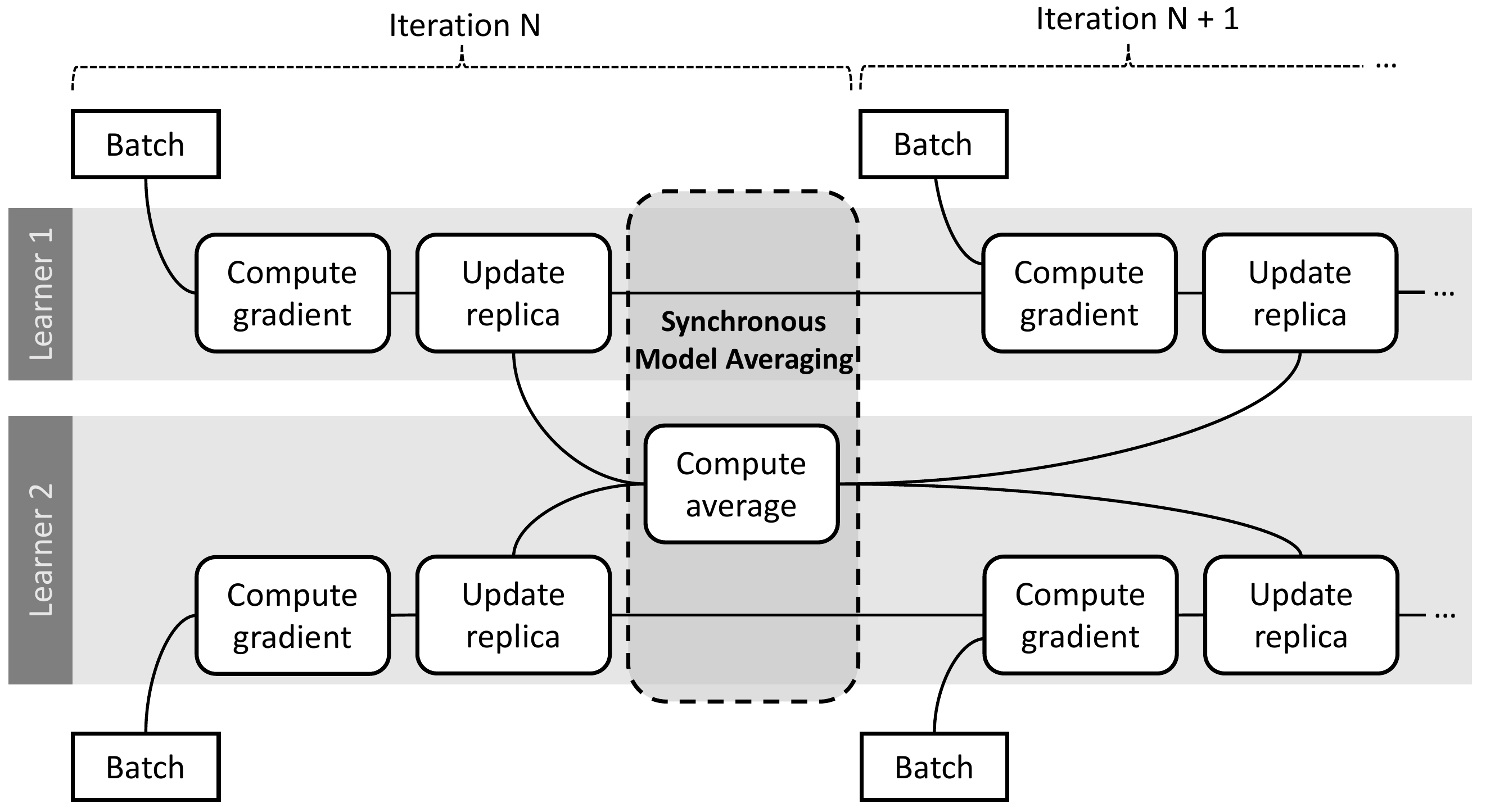}
  \caption{Parallel training with two learners \captext{Each learner
    independently trains a model replica. The replica is updated based on the
    locally computed gradients as well as corrections derived through model
    averaging.}}\label{fig:sma_overview}
\end{figure}

\section{Synchronous Model Averaging with Learners}
\label{sec:sma}

Our approach relies on the concept of a \emph{learner}, which independently
trains a model replica for a given input batch~(\Cref{sec:learner}). Having
many independent learners requires careful synchronisation in order to achieve
high statistical efficiency. We introduce a new algorithm, named
\emph{synchronous model averaging}~(\sync), that consolidates the model updates
computed by many learners~(\Cref{sec:algorithm_design}). After that, we discuss
how to train multiple learners per GPU~(\Cref{sec:scalable_synchronisation})
and how to determine the number of learners to use~(\Cref{sec:autotuning}).

\subsection{Independent learners}
\label{sec:learner}

Parallel S-SGD imposes tight synchronisation when processing partitioned
batches. The gradients computed based on \emph{all} model replicas are
aggregated, and the obtained result is incorporated by all replicas. After each
iteration, before the computation of gradients for the next batch begins, all
replicas are therefore the same.

Our idea is to introduce more diversity into the learning process based on the
notion of a \emph{learner}. A learner is an entity that trains a single model
replica \emph{independently} with a given batch size. \Cref{fig:sma_overview}
shows two learners, each being assigned a different complete batch. A learner
computes a gradient and immediately updates its replica based on the
gradient. It then continues with the gradient computation for the next
batch. As we describe in the next section, to prevent learners from diverging,
each learner also applies a \emph{correction} to its model, which is
incorporated synchronously as part of the next update of the replica.

In contrast with parallel S-SGD, model replicas with learners evolve
independently because they are not reset to a single global model after each
batch. Unlike asynchronous learning approaches~\cite{Recht2011, Noel2014, Zhang2014b},
each replica is corrected in each iteration to maintain the convergence of the
learning process. Learners enable us to achieve both high statistical
efficiency and hardware efficiency, avoiding the trade-off between the two
faced by other parallel training approaches.

\begin{figure}[tb]
  \centering
  \includegraphics[scale=0.4]{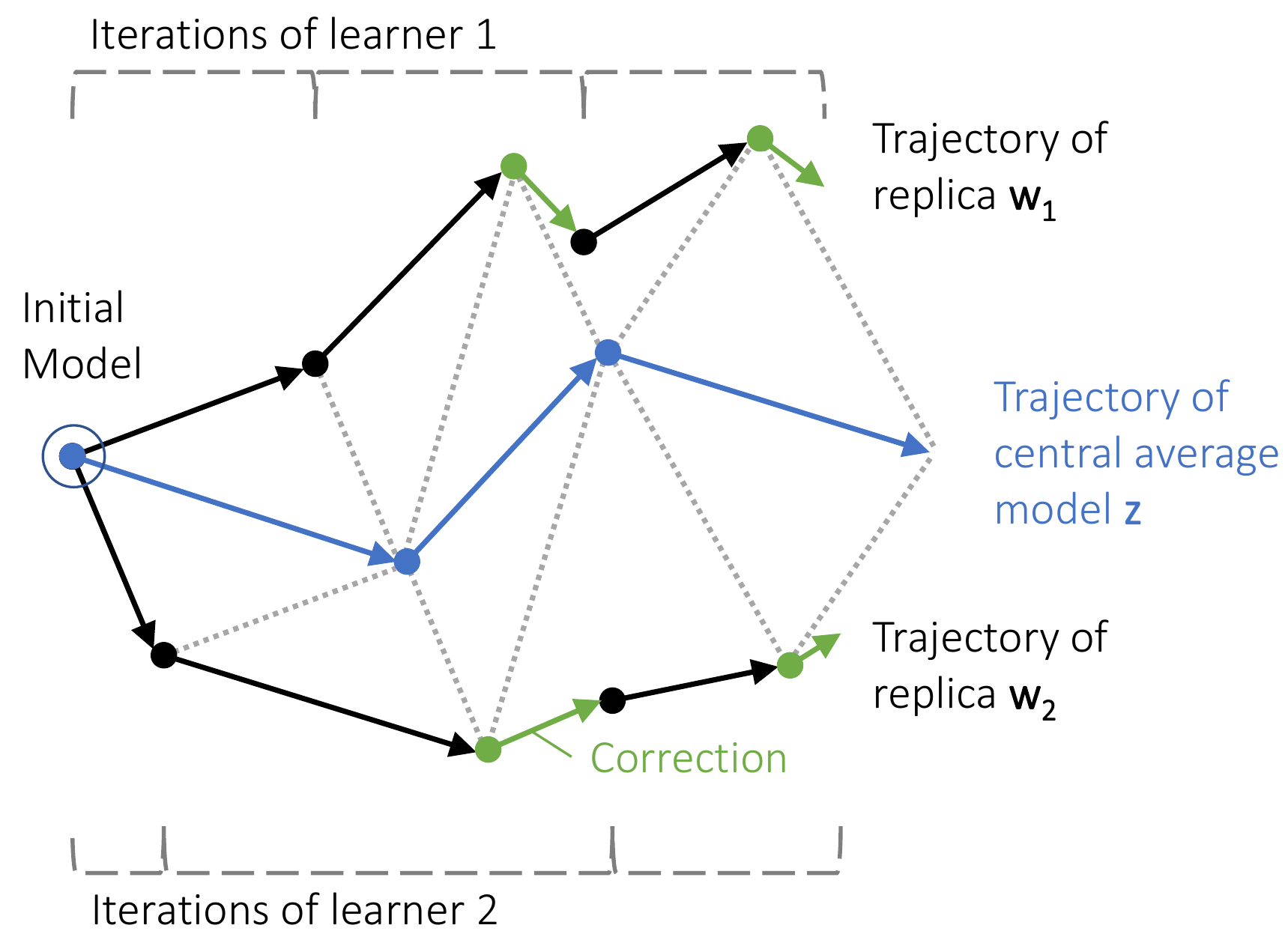}
  \caption{Intuition behind \sync \captext{Two replicas
    $\mathsf{w}_{1}$ and $\mathsf{w}_{2}$ are trained by two learners. Their
    respective model updates are incorporated in a central average
    model~$\mathsf{z}$. Based on the latter, corrections for $\mathsf{w}_{1}$
    and $\mathsf{w}_{2}$ are derived and applied after each
    iteration.}}\label{fig:convergence_momentum}
\end{figure}

\subsection{\Sync algorithm}
\label{sec:algorithm_design}

To synchronise the local models of learners, we propose \emph{synchronous model
  averaging}~(\sync), a new algorithm based on model
averaging~\cite{Polyak1992, Polyak1990, Ruppert1988}. \sync consolidates the
model updates of learners by maintaining a \emph{central average model}. We
illustrate the idea behind \sync in \Cref{fig:convergence_momentum}. Starting
with the initial model, two learners train replicas, $\mathsf{w}_{1}$ and
$\mathsf{w}_{2}$, with distinct batches. Once the learners have computed the
gradients and updated their local replicas, the updates are applied to a
central average model. The average model is used to compute corrections for
each local replica, ensuring that they follow the trajectory of the central
average model.

\emph{Corrections} are defined as the difference of a replica with the current
central average model. A correction represents a penalty of the replica
disagreeing with the consensus in the previous iteration~\cite{Boyd2011}, and
pulls a replica toward the average value over all replicas. The expectation is
that all replicas gradually agree on the optimality of the central average
model, and eventually reach the same minima.

In general, a desirable property of model averaging is that the asymptotic
variance of the central average model is lower than the variance of individual
model replicas. This difference prevents the accumulation of training errors on
each learner~\cite{Bottou2018}. As the number of learners increases, however,
the difference in variance tends to decrease, preventing the central averaging
model from converging faster than the learners. To compensate for this effect,
\sync uses \emph{momentum}, as introduced in \Cref{sec:ml-background}. Unlike
past work that applies momentum for individual model replicas
only~\cite{Zhang2015,Li2014b}, we incorporate the directions of model weights
into the updates to the central average model: updates in directions of
persistent descent are kept; those in other directions are cancelled or
diminished.
We use Polyak's momentum method~\cite{Polyak64momentum} because, compared to
Nesterov's accelerated gradient~\cite{Nesterov1983}, it fits model averaging
better: the update to the central average model is computed by all learners
based on their current position and not an estimated one~\cite{Sutskever2013}.

\begin{algorithm}[t]
	\caption{\textbf{\textsf{Synchronous model averaging~(SMA)}}}\label{alg:sma}
	\SetKwInOut{Input}{input}%
	\SetKwInOut{Output}{output}%
	\Input{
		$\mathsf{w}_{0}$, an initial model;\\
		\ $\mathsf{w}_{1}\ldots \mathsf{w}_{k}$, $k$ model replicas trained by $k$~learners;\\
		\ $\mathbb{B}$, a set of batches;\\
		\ $\upgamma$, a learning rate parameter;\\
		\ $\upmu$, a momentum parameter;
	}
	\Output{
		$\mathsf{z}$, the trained model.
	}
	\BlankLine
	\tcp{{\smaller Initialise central average model and its previous version}}
	$\mathsf{z}\leftarrow \mathsf{w}_{0}$ \label{alg:sma:init_1} \; 
	$\mathsf{z}_{\mathit{prev}} \leftarrow \emptyset$ \label{alg:sma:init_2} \;

	\BlankLine
	\While{target accuracy not reached $\land$ $|\mathbb{B}|\geq k$}{
    \label{alg:sma:loop_1}
    \tcp{{\smaller $i$-th iteration of the learning algorithm}}
		
		$c_{1},\ldots, c_{k} \leftarrow \emptyset,\ldots,\emptyset$ \;
		\For{$j \in \{1,\ldots,k \}$}{
      \tcp{{\smaller $j$-th learner in the $i$-th algorithm iteration}}
			$B_j \leftarrow select(\mathbb{B})$ \label{alg:sma:select} 
			\tcp*{{\smaller Select batch 
			for learner $j$}}
			$\mathbb{B} \leftarrow \mathbb{B} \setminus \{B_j\}$
			\tcp*{{\smaller Remove the batch}}
			$g_{j} \leftarrow \upgamma \nabla \ell_{B_j}(\mathsf{w}_{j})$
      \label{alg:sma:gradient}
			\tcp*{{\smaller Gradient for replica $j$}}
			$c_{j} \leftarrow \upalpha (\mathsf{w}_{j} - \mathsf{z})$
      \label{alg:sma:correction}
			\tcp*{{\smaller Correction for replica $j$}}
			$\mathsf{w}_{j} \leftarrow 
			\mathsf{w}_{j} - g_{j} - c_{j}$
      \label{alg:sma:update}
			\tcp*{{\smaller Update replica $j$}}
		}
		\tcp{{\smaller Update central average model}}
		$\mathsf{z}'\leftarrow \mathsf{z}$ \;
		$\mathsf{z} \leftarrow \mathsf{z} + \sum_{j=1}^{k} 
		c_{j} + \upmu (\mathsf{z} - \mathsf{z}_{\mathit{prev}})$
        \label{alg:sma:central}
    \;

		$\mathsf{z}_{\mathit{prev}}\leftarrow \mathsf{z}'$\;
		
    \label{alg:sma:loop_2}}
\end{algorithm}

We formalise the SMA algorithm in \autoref{alg:sma}. It takes as input a model,
initialised as $\mathsf{w}_{0}$, a set of $k$~model replicas
$\mathsf{w}_{1}\ldots \mathsf{w}_{k}$ that are managed by $k$~learners, a set
of batches~$\mathbb{B}$, along with two hyper-parameters: the learning
rate~$\upgamma$ and the momentum~$\upmu$. Upon termination, the algorithm
returns the trained model.

First \sync initialises the central average model~$\mathsf{z}$ and a reference
to a previous version of it, $\mathsf{z}_{\mathit{prev}}$
(lines~\ref{alg:sma:init_1}--\ref{alg:sma:init_2}). It defines an iterative
learning process (lines~\ref{alg:sma:loop_1}--\ref{alg:sma:loop_2}) that
terminates when the target accuracy has been reached by the central average
model~$\mathsf{z}$ or there are insufficient batches available.

In each iteration, a learner~$j$ proceeds as follows: (i)~it selects a
batch~$B_j$ (\autoref{alg:sma:select}) and, using the batch and its
replica~$\mathsf{w}_{j}$, the learner computes a gradient~$g_j$ (as in
\autoref{eq:iter}) under the given learning rate~(\autoref{alg:sma:gradient});
(ii)~it computes a correction~$c_j$ as the difference between the
replica~$\mathsf{w}_{j}$ and the central average model~$\mathsf{z}$ where
$\upalpha \approx 1/k$ is a constant~(\autoref{alg:sma:correction}); and
(iii)~the model replica~$\mathsf{w}_{j}$ is then updated by applying the
gradient~$g_{j}$ and the correction~$c_{j}$ (\autoref{alg:sma:update}).

The iteration completes with an update to the central average
model~$\mathsf{z}$~(\autoref{alg:sma:central}). This update is twofold: (i)~it
includes the corrections derived for all the $k$~model replicas assigned to the
independent learners, which represent the current differences of the replicas
with respect to~$\mathsf{z}$; and (ii)~a second component exploits the
configured momentum and the previous version of the central average
model~$\mathsf{z}_{\mathit{prev}}$ to accelerate convergence by maintaining the
direction of gradients~\cite{Hinton2012}. In the $i$-th iteration of \sync,
$\mathsf{z}_{\mathit{prev}}$ is the model at the beginning of the
$(i$$-$$1)$-th iteration.

Similar to most parallel training approaches~\cite{DBLP:journals/corr/Krizhevsky14,
  Zhang2015}, \sync benefits from an online adaptation of
hyper-parameters. Based on the accuracy observed after each iteration, the
learning rate (parameter~$\upgamma$ in \autoref{alg:sma}) may be reduced
step-wise to overcome oscillation of the trained model and improve
accuracy~\cite{LeCun2012}. Such adaptation can be realised by
updating~$\upgamma$ directly in each iteration in \sync. For example, when
training ResNet-50, it is common to reduce the learning rate twice, at the 30th
and 60th epochs~\cite{HeZRS15}, which are chosen empirically based on the
accuracy observed after each iteration.

With \sync, however, oscillatory behaviour is observed on the
central average model, whereas a change in the learning rate affects the 
training of each model separately. Since \sync does not reset all models in each
iteration, the accuracy may not necessarily improve each time the learning rate
changes. When detecting such a situation, we therefore restart \sync:
\autoref{alg:sma} is executed again with the latest version of the central
average model~$\mathsf{z}$ as the new initial model~$\mathsf{w}_{0}$.

\begin{figure}[tb]
  \centering \includegraphics[scale=0.34]{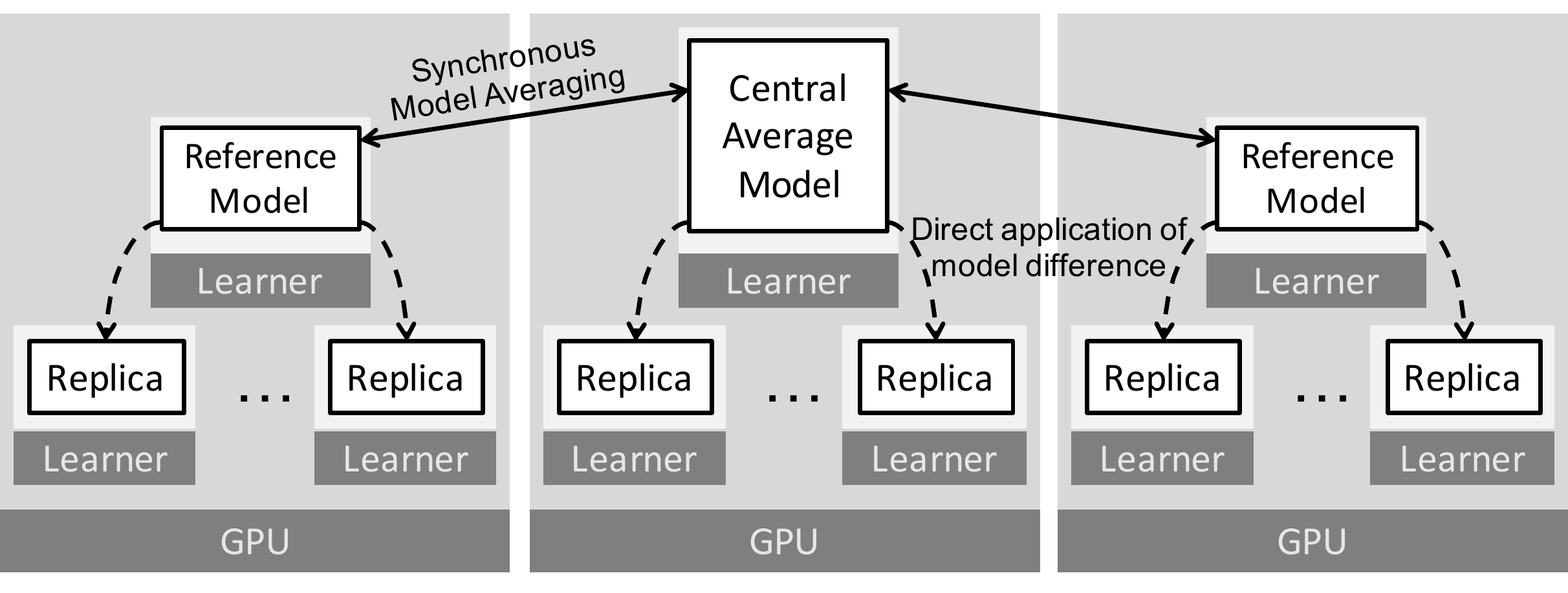}
  \caption{Synchronising multiple learners per GPU \captext{\sync is
    applied for one reference model per GPU; further learners on the GPU
    incorporate differences of their replicas and the reference model.}
      }\label{fig:multiple_learners}
\end{figure}

\subsection{Training multiple learners per GPU}
\label{sec:scalable_synchronisation}

When selecting a small batch size for training in order to achieve high
statistical efficiency, training a single model replica may not saturate all
resources of a GPU. Learners enable us to decouple the processing of a batch
from the available hardware though. To achieve high hardware utilisation, we
can execute multiple learners per GPU. If a batch size as low as 2 is used,
dozens of learners can run concurrently on a multi-GPU server. 

Given a potentially large number of learners, we take advantage of the fact 
that some learners reside on the same GPU and access shared memory, which is at 
least an order of magnitude faster than PCIe for inter-GPU communication.   
Rather than aggregating all differences in a single step, we therefore separate 
synchronisation on the intra-GPU and inter-GPU level.

We organise synchronisation as illustrated in
\Cref{fig:multiple_learners}. To synchronise the learners executing on a single
GPU, one learner is chosen to manage a \emph{reference model}. Each learner
then computes the difference between its model replica and the local reference
model. This difference is then applied to the respective replica. At the
global level, the \sync algorithm is executed. It uses one of the local
reference models as the central average model ($\mathsf{z}$ in
\autoref{alg:sma}); all other reference models (one per GPU) are the replicas
incorporated into the model averaging process ($\mathsf{w}_{j}$ in
\autoref{alg:sma}).

\subsection{Choosing the number of learners}
\label{sec:autotuning}

\begin{algorithm}[t]
	\caption{\textbf{\textsf{Selecting the number of 
	learners per GPU}}}\label{alg:learners}
	\SetKwInOut{Input}{input}%
	\SetKwInOut{Output}{output}%
	\Input{ $\tau$, a throughput threshold parameter;
	}
	\BlankLine
	${l}_{1}, \ldots, l_{m} \leftarrow 1,\ldots, 1$
	\label{alg:learners:init}
	\tcp*{{\smaller Number of learners for each of $m$~GPUs}}
	
	${t}'_{1}, \ldots, t'_{m} \leftarrow 0,\ldots, 0$
	\tcp*{{\smaller Throughput observed earlier for the $m$~GPUs}}
	\BlankLine
	\While{{\textsf{\scriptsize{}SMA}} executes}{
		\For{$g\in \{1,\ldots, m\}$}{
			\label{alg:learners:for_1}
			\tcp{{\smaller Observe learning throughput of $g$-th GPU}}
			${t} \leftarrow $ \emph{get-current-throughput}$(g)$\;
			\tcp{{\smaller Adapt number of learners for $g$-th GPU}}
			\lIf{$t - t'_g > \tau$}{
				${l}_{g} \leftarrow {l}_{g} + 1$
				\label{alg:learners:add}
			}
			\lElseIf{$t < t'_g \ \wedge l_g > 0$}{
				${l}_{g} \leftarrow {l}_{g} - 1$
				\label{alg:learners:rm}
			}
			${t}'_{g} \leftarrow t$
			\label{alg:learners:for_2}\;
		}
	}
\end{algorithm}

The number of learners per GPU or, put differently, the number of model
replicas to be trained in parallel, is an important parameter. It must be
chosen carefully for a given batch size: when training too few replicas, a \gpu
is under-utilised, wasting resources; when training too many, the execution of
otherwise independent learners is partially sequentialised on a \gpu, leading
to a slow-down.

We propose to tune the number of learners per GPU based on the training
throughput at runtime. By observing the number of processed batches per second,
we can detect under- and over-utilisation of a \gpu during training.  As
described in \autoref{alg:learners}, we initially use a single learner per
GPU~(\autoref{alg:learners:init}). For each GPU, we then consider the learning
throughput (lines~\ref{alg:learners:for_1}--\ref{alg:learners:for_2}): if a
significant increase in throughput is observed, \ie the increase is a above a
predefined tolerance threshold~$\tau$, a new learner is added to the respective
GPU~(\autoref{alg:learners:add}). Upon observing a decrease in throughput, we
reduce the number of learners again~(\autoref{alg:learners:rm}).

Changing the number of learners is also beneficial in terms of statistical
efficiency. Having initially few learners reduces the noise of stochastic
gradients, fostering convergence of the reference models and thus the central
average model. Eventually, though, this hampers the optimality of convergence
as a smaller part of the loss space is explored.\footnote{Similar observations
  have been made regarding dynamic changes of the batch size~\cite{Smith2017}
  and learning rates~\cite{GoyalDGNWKTJH17}.} By increasing the parallelism
gradually, we avoid this issue. Intuitively, 
a small initial number of learners allows the central average model to reach
the neighbourhood of the solution quickly, which is then comprehensively
explored with increased training parallelism.

\section{Crossbow System Design}
\label{sec:system}

To support the training of deep learning models using \sync, the design of \sys
has several unique features:

\tinyskip
  
\noindent
(1)~Since we train multiple learners per \gpu, \sys must share GPUs
efficiently. \sys executes learners concurrently on a GPU by scheduling each to
run on a separate \gpu stream.

\tinyskip
  
\noindent
(2)~We decide on the number of learners per GPU at runtime. The design of
\sys must support changing the number of learners per GPU dynamically based on 
the available GPU resources.

\tinyskip
  
\noindent
(3)~\sync synchronises all learners when they access the central average
model. The design of \sys must implement this global synchronisation operation
efficiently, and exploit concurrency during synchronisation to avoid bottlenecks.

\tinyskip

Next we introduce the main component of \sys{}'s design in
\Cref{sec:system-overview}. Based on the above requirements, we then describe
hows tasks execute~(\Cref{sec:task-execution-model}) and are
scheduled~(\Cref{sec:scheduling}), and how the number of learners is tuned
dynamically~(\Cref{sec:learner_tuning}). We finish with an explanation of
memory management in \Cref{sec:memory_mgmt}.

\subsection{System overview}
\label{sec:system-overview}

\begin{figure}[t]
\centering
\includegraphics[width=0.95\columnwidth]{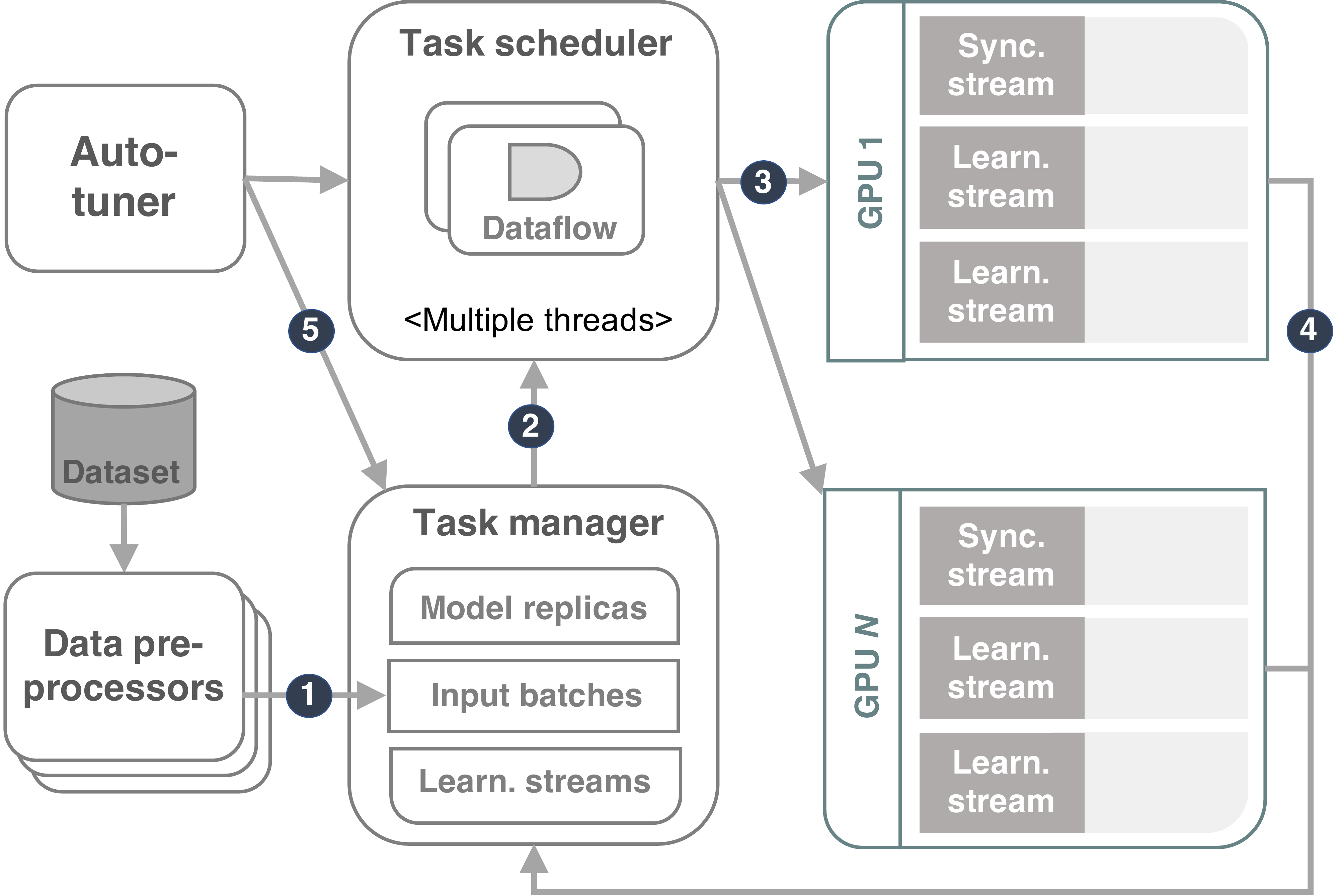}
\caption{\sys design}\label{fig:overview}
\end{figure}

\Cref{fig:overview} shows the main components of \sys:

\tinyskip

\noindent
(1)~The \textbf{data pre-processors} read the training dataset into memory and
arrange samples into batches, possibly after some transformations such as image
decoding and cropping.

\tinyskip

\noindent
(2)~The \textbf{task manager} controls the pools of model replicas, input
batches and learner streams. It handles task completion events originating from
the GPUs.

\tinyskip

\noindent
(3)~The \textbf{task scheduler} assigns learning tasks to GPUs based on the
available resources. It also triggers synchronisation operations at the end of
each iteration.

\tinyskip

\noindent
(4)~The \textbf{auto-tuner} monitors the training throughput and creates new
learners on a GPU on-demand.

\tinyskip

\Cref{fig:overview} shows how \sys executes an iteration of \sync: the data
pre-processors populate the input batch pool with pointers to data, one
complete batch at a time~(step~\myc{1}). The task scheduler checks if a model
replica and a co-located learner stream are available (step~\myc{2}) and then
schedules a learning task~(step~\myc{3}). Upon completion, the task manager
handles the event and returns the learner stream and the model replica to the
pool~(step~\myc{4}). It also frees up a slot in the input batch pool, to be
populated by one of the data pre-processors~(step~\myc{1}). The auto-tuner
monitors the rate at which learning tasks complete and interrupts the training
by adding a new learner~(step~\myc{5}).

\subsection{Task execution}
\label{sec:task-execution-model}

\begin{figure}[t]
\centering
\includegraphics[width=0.95\columnwidth]{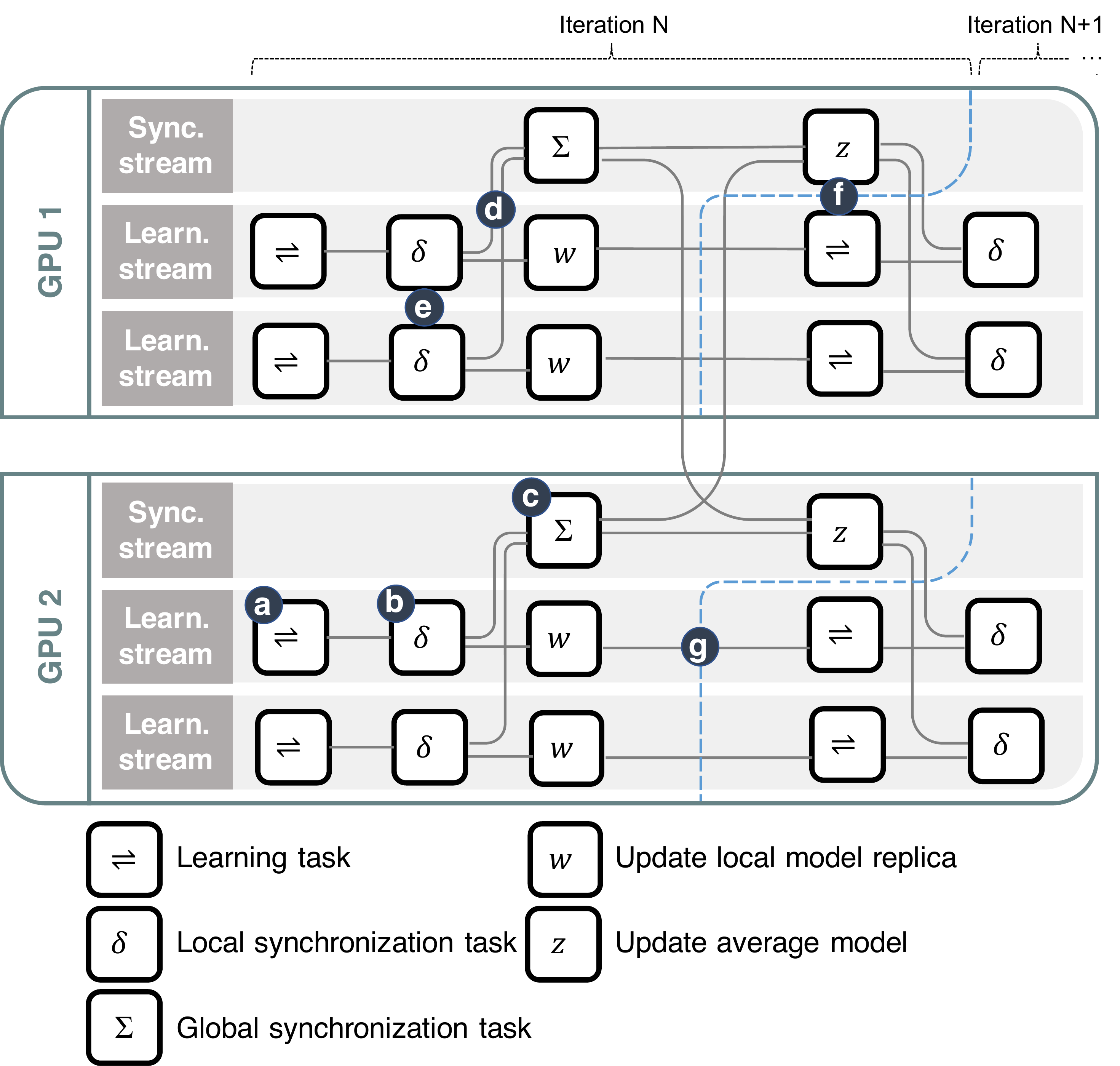}
\caption{\mbox{Dataflow graph on 2~GPUs with 2~learners each}
  \captext{The figure shows how tasks are scheduled on streams during two
  successive iterations of \sync, separated by a dashed line; solid lines are
  data dependencies between tasks.}}
  \label{fig:dependencies}
\end{figure}
  
\sys trains a deep learning model by executing a dataflow graph, as shown in
\Cref{fig:dependencies}. The dataflow graph consists of a set of \emph{learning
  tasks} interleaved with \emph{synchronisation tasks}. \sys represents the
processing layers of a deep learning model as a graph of operators (\eg a
convolution or a matrix multiplication). A \emph{learning task} encapsules
multiple operators (see~\myc{a}). It takes as input a batch of training samples
and a model replica and outputs a gradient. The task scheduler executes a
learning task on any of the \emph{learner streams} available on the \gpu on
which the replica used by that learning task resides.

Synchronisation tasks can be local or global: (i)~a \emph{local synchronisation
  task} computes the difference between a model replica and the central average
model~(see~\myc{b}). It uses the difference, together with the gradient of the
corresponding learning task, to update the model replica. There is one local
synchronisation tasks per replica on the same learner stream as the
corresponding learning task; (ii)~a \emph{global synchronisation task}
aggregates all local differences to update the central average
model~(see~\myc{c}). \sys allocates one average model replica per \gpu, and
uses a separate stream, the \emph{synchronisation stream}, to execute global
synchronisation tasks.

Placing a global execution barrier between the synchronisation and all
preceding learning tasks would be expensive: it would block the task scheduler
and delay the GPUs already waiting for tasks to complete. Instead, the task
scheduler overlaps the synchronisation tasks from one iteration with the
learning tasks of the next.

When overlapping tasks, the task scheduler considers \emph{data dependencies}
between tasks~(see~\myc{d}). The horizontal direction in the figure represents
time, and the vertical direction a spatial GPU partitioning based on streams. A
line connecting two tasks denotes that the task on the right cannot start until
the one on the left has completed.

Within an iteration, all local synchronisation tasks can execute concurrently
(see~\myc{e}). They only require read-only access to the central average model
in order to compute the difference from each model replica. The local
synchronisation tasks, however, depend on the average model being updated
consistently on each GPU by the global synchronisation tasks of the previous
iteration.

Global synchronisation tasks can also execute concurrently within an iteration
(see~\Cref{sec:scalable_synchronisation}). The intra-GPU operations of a global
synchronisation task execute as soon as their dependency to a local
synchronisation task is satisfied (see~\myc{d}). The inter-GPU operations are
implemented as a collective \emph{all-reduce} primitive~\cite{sergeev2018horovod}.
All-reduce creates a ring topology in which each GPU exchanges data partitions
with its peers. A GPU reduces the partition data that it receives by combining it
with its own, and eventually every GPU holds the final aggregated data. 
As a result, all-reduce evenly distributes the computation of the update for
the average model among the GPUs.

The global synchronisation tasks in one iteration can execute concurrently with
the learning tasks in the next (see~\myc{f}). Once the local synchronisation
tasks have completed, each replica is updated independently. At this point, the
task scheduler can issue the next learning task to the learner stream without
waiting for other tasks to complete (see~\myc{g}).

\subsection{Task scheduling}
\label{sec:scheduling}

In each iteration of \sync, the task scheduler schedules one learning task for
each model replica in the pool, followed by synchronisation tasks. As the task
manager returns newly-updated model replicas to the pool, the task scheduler
schedules further learning tasks and associates the next batch with a model
replica on a first-come, first-served basis. Compared to round-robin
scheduling, as used by PyTorch~\cite{pytorch} or
TensorFlow~\cite{Abadi2016osdi}, this improves hardware efficiency because the
task scheduler does not need to wait for a given replica to become
available. After an assignment, the task scheduler hands the learning task over
to one of its worker threads, in particular, one that runs on the same socket
as the task's designated GPU. The worker thread issues the task to one of the
GPU's streams as a sequence of kernels. All kernel calls to the GPU are
non-blocking, and the thread returns immediately to schedule the next task.

A challenge for the task scheduler is to ensure that tasks that can run
concurrently are executed concurrently by a GPU. As there are multiple
streaming multi-processors~(SMs) per GPU with no shared resources among them, a
\gpu can execute multiple tasks concurrently by assigning them to different
sets of SMs. This favours our approach of training with small batch sizes
because the kernels of a learning task usually require few SMs. As a solution,
the task scheduler assigns concurrent tasks to different GPU streams, which
enables the GPU to make efficient internal scheduling decisions: tasks
submitted to the same stream execute in issue order; tasks on different streams
may execute concurrently. A worker thread issues the task's kernels to its
assigned streams, along with any event generators or handlers.

The task scheduler uses GPU events to respect any data dependencies between
submitted tasks. If there is a dependency between tasks~$\uptau_{1}$ and
$\uptau_{2}$, the task scheduler submits an \emph{event generator} after
$\uptau_{1}$ completes and an \emph{event handler} before $\uptau_{2}$
begins. When the event generator on $\uptau_{1}$'s stream executes, it signals
$\uptau_{2}$'s handler that all preceding tasks on that stream, $\uptau_{1}$
included, have completed; when the event handler on $\uptau_{2}$'s stream
executes, all subsequent tasks on that stream block until it receives a signal
from $\uptau_{1}$'s event generator. Analogous to the task scheduler, the task
manager also uses multiple threads to handle task completion events in
parallel, returning model replicas return to the pool in a timely manner.

\subsection{Tuning learners}
\label{sec:learner_tuning}

The \emph{auto-tuner} changes the number of learners per GPU at runtime without
prior knowledge of the trained model (\eg the number and computational
complexity of each model layer) or the training environment (\eg the number and
capability of each GPU). To this end, it implements the adaptation procedure
introduced in \Cref{sec:autotuning} and formalised in \autoref{alg:learners}.

The auto-tuner measures the training throughput by considering the rate at
which learning tasks complete, as recorded by the task manager. As defined in
\autoref{alg:learners}, the number of learners per GPU is then increased or
decreased based on the observed change in training throughput. Note that, on a
server with homogeneous GPUs, the auto-tuner may measure only the throughput of
a single GPU to adapt the number of learners for all GPUs.

The auto-tuner initiates the creation of a new learner after the learning and
synchronisation tasks of the current iteration have been scheduled. Adding a
learner to a GPU requires allocating a new model replica and a corresponding
learner stream.
The auto-tuner places temporarily a global execution barrier between two
successive iterations (step~\myc{f} in~\Cref{fig:dependencies}), avoiding
overlap with other tasks. The new model replica is initialised with the latest
value of the average model. The auto-tuner also locks the resources pools,
preventing access by the task scheduler or manager during resizing.

Even for large models, such as ResNet-50, auto-tuning completes within
milliseconds. The main overhead comes from the memory allocation and the
initialisation of the model weights. Since model weights and their gradients
are kept in contiguous memory, a single allocation call suffices.

\begin{figure}[tb]
\def\arraystretch{1.1}
\centering
\small
\begin{tabularx}{0.98\columnwidth}{@{}llrr>{\raggedleft}X@{}}
\toprule
{Model} & 
Dataset & 
{Input size (MB)} & 
{\# Ops} &
{Model size (MB)} 
\tabularnewline
\midrule
\textsf{LeNet} & 
MNIST & 
179.45 & 
24 & 
4.24
\tabularnewline

\textsf{ResNet-32} & 
CIFAR-10 & 
703.12 & 
267 & 
1.79
\tabularnewline

\textsf{VGG-16} & 
CIFAR-100 & 
703.12 & 
121 & 
57.37 
\tabularnewline

\textsf{ResNet-50} & 
ILSVRC~2012 & 
1,073,375.25 & 
384 & 
97.49
\tabularnewline

\bottomrule

\end{tabularx}
\captionof{table}{Deep learning benchmarks and datasets used}
\label{tab:apps}
\end{figure}

\subsection{Memory management}
\label{sec:memory_mgmt}

Data pre-processors transfer the input data from CPU to GPU memory using direct
memory access~(DMA). They write the pre-processed training samples to a
page-aligned, page-locked circular buffer whose memory range is registered to
the GPU's address space. This memory can be read directly by the GPU with
higher bandwidth than unregistered, pageable memory. The buffer size must
accommodate at least one input batch per learner (\ie enough for a complete
iteration of \sync). \sys uses double buffering to create a pipeline between
data pre-processors and the task scheduler. When the pre-processors stall the
pipeline because it takes more time to prepare the data on the CPU than to
process it on a GPU, some or all of the input data transformations are
scheduled on the GPUs as a preamble to each learning task.

Deep learning models require more memory to store the output of their dataflow
operators than the model itself. For example, the ResNet-50 model is
97.5\unit{MB} in size but consumes 7.5\unit{GB} of memory to store the outputs
from 384~operators. The memory requirement scales linearly with the batch size,
as well as the number of learners. \sys must therefore reduce the memory
footprint of each learner when training multiple of them per GPU.

\sys, similar to TensorFlow~\cite{Abadi2016osdi},
MxNet~\cite{DBLP:journals/corr/ChenLLLWWXXZZ15} and
Super-Neurons~\cite{Wang18}, devises an \emph{offline memory plan} to reuse the
output buffers of operators using reference counters. During initialisation,
\sys traverses the operators of a learning task.
When visiting an operator, it considers preceding operators and reuses an
output buffer if the reference count is zero; otherwise it assumes that a new
output buffer must be created. To account for data dependencies, it then
decrements the reference counter of the operator's inputs and increments the
counters of the operator's output.
Such an offline plan reduces the memory footprint of a learner by up to 50\%
because outputs are mostly reused during the backwards phase.
  
When executing multiple learners per GPU, however, replicating the offline plan
for each learner would over-allocate memory. \sys exploits that, in practice,
not all instances of the same operator would execute concurrently. This enables
the sharing of some of the output buffers among learners on the same GPU using
an \emph{online memory plan}. For each operator, the task scheduler maintains a
pool of output buffer pointers to GPU memory.
Pools are shared by all learners on the same GPU. At runtime,
when the task scheduler considers an operator for execution, it reuses the
first available buffer in the output buffer pool; if none are available, it
allocates a new one. The task scheduler increments the reference counter of the
operator's output according to its data dependencies and issues the kernel on a
stream. When the operator completes, the task manager decrements the reference
counter of the operator's input and output buffers. A buffer with zero
references returns to the pool, and it can be reused by other learning tasks.

\begin{figure}[tb]
\centering
\begin{subfigure}[t]{0.23\textwidth}
\centering
\includegraphics[width=\textwidth]{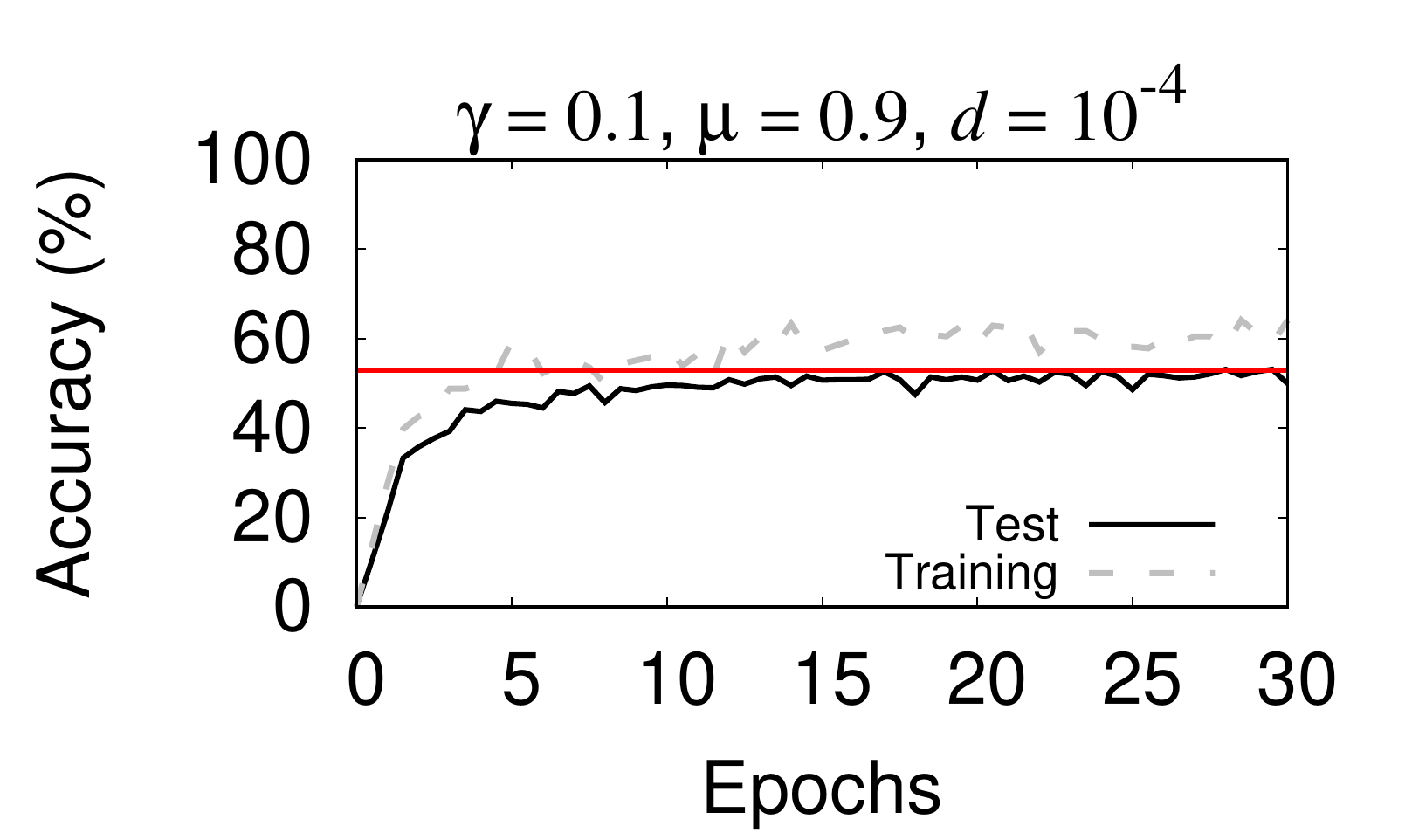}
\caption{ResNet-50}
\label{fig:resnet50-setup}
\end{subfigure}
~
\begin{subfigure}[t]{0.23\textwidth}
\includegraphics[width=\textwidth]{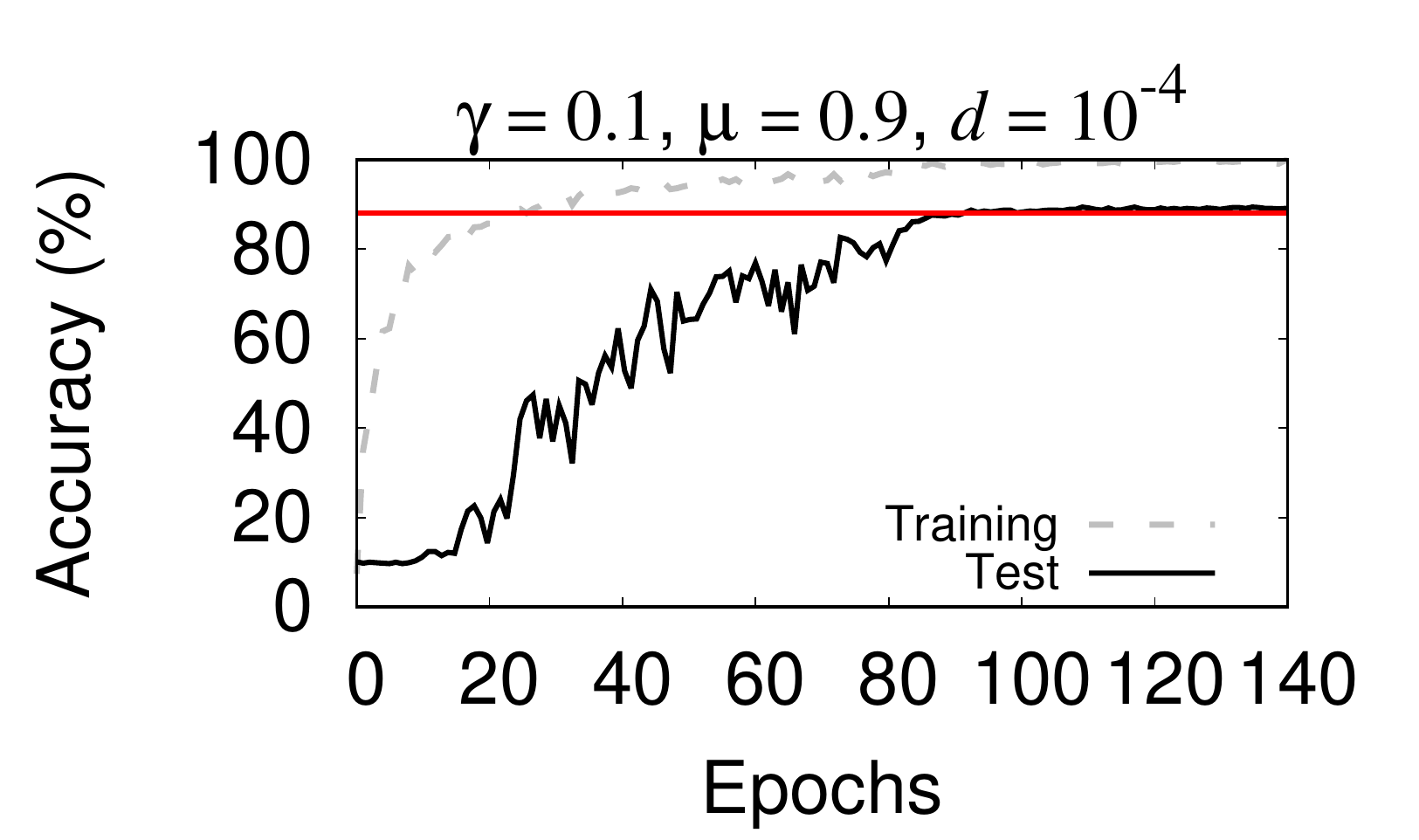}
\caption{ResNet-32}
\label{fig:resnet32-setup}
\end{subfigure}
\\
\begin{subfigure}[t]{0.23\textwidth}
\includegraphics[width=\textwidth]{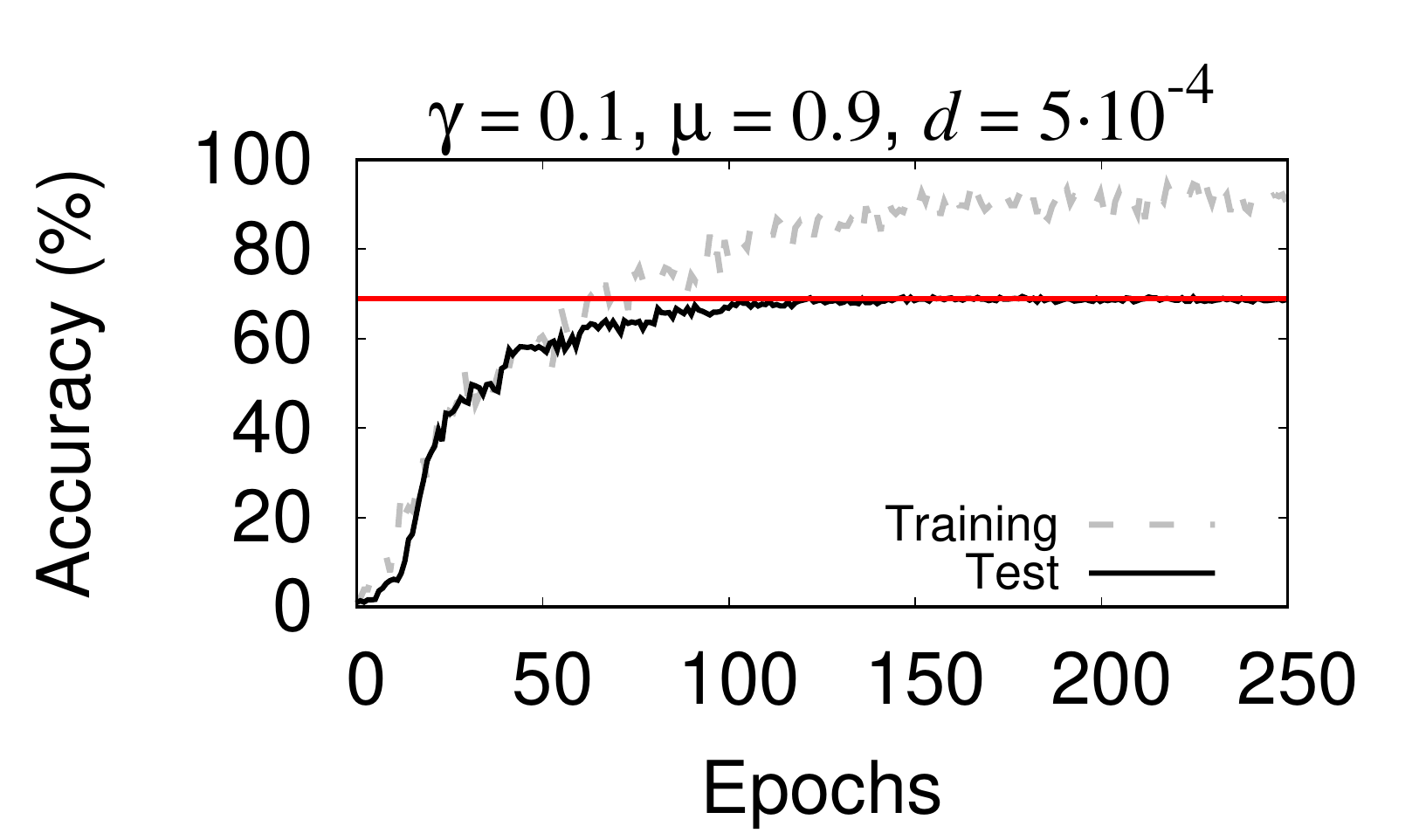}
\caption{VGG}
\label{fig:vgg-setup}
\end{subfigure}
\begin{subfigure}[t]{0.23\textwidth}
\includegraphics[width=\textwidth]{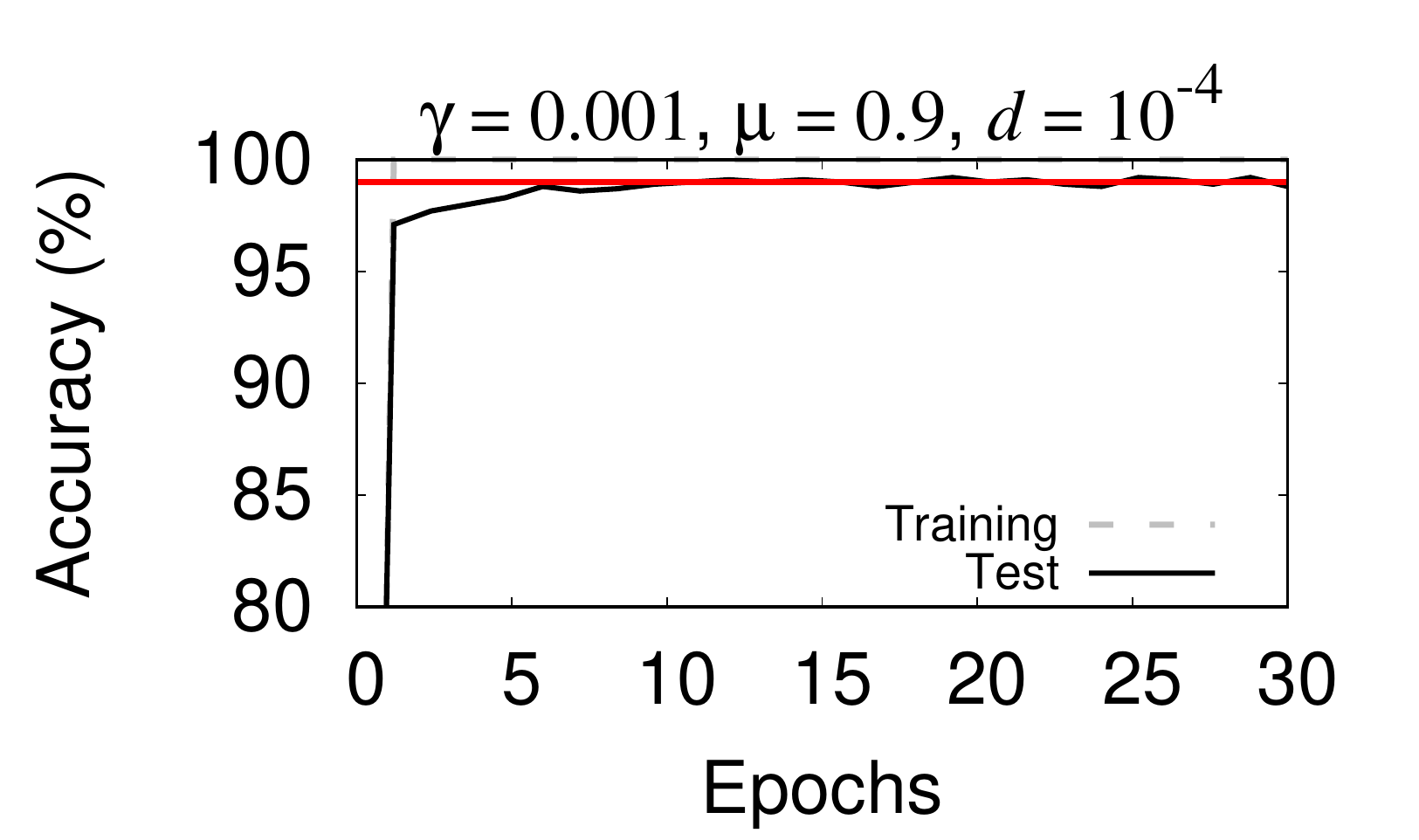}
\caption{LeNet}
\label{fig:lenet-setup}
\end{subfigure}
\caption{TensorFlow's convergence over epochs \captext{Test and training
    accuracy over epochs for a given learning rate $\upgamma$, momentum
    $\upmu$, and weight decay $d$. Red lines indicate our test accuracy targets.}}
\label{fig:acc-target}
\end{figure}

\section{Evaluation}
\label{sec:evaluation}

In this section, we evaluate the performance of our \sys prototype when
training on a multi-GPU server. We begin by comparing its behaviour against
TensorFlow~\cite{Abadi2016osdi} using four deep learning macro-benchmarks, as
we vary the number of GPUs and the number of learners per
GPU~(\Cref{sec:eval:scaling}). Next, we assess the impact of \sys's core
features through a set of micro-benchmarks: we explore the benefits of
executing multiple learners per GPU in terms of statistical and hardware
efficiency~(\Cref{sec:eval:batch-size}); and we measure the ability of the
auto-tuning mechanism to identify the best number of learners per
GPU~(\Cref{sec:eval:auto-tuning}) and the impact of the model averaging used in
\sync~(\Cref{sec:eval:sma}). Finally, we quantify the efficiency of the
synchronisation task implementation~(\Cref{sec:eval:overhead}).

\subsection{Experimental set-up}
\label{sec:eval:settings}

\begin{figure*}[tb]
  \centering
  \begin{subfigure}[t]{0.325\textwidth}
    \includegraphics[width=\textwidth]{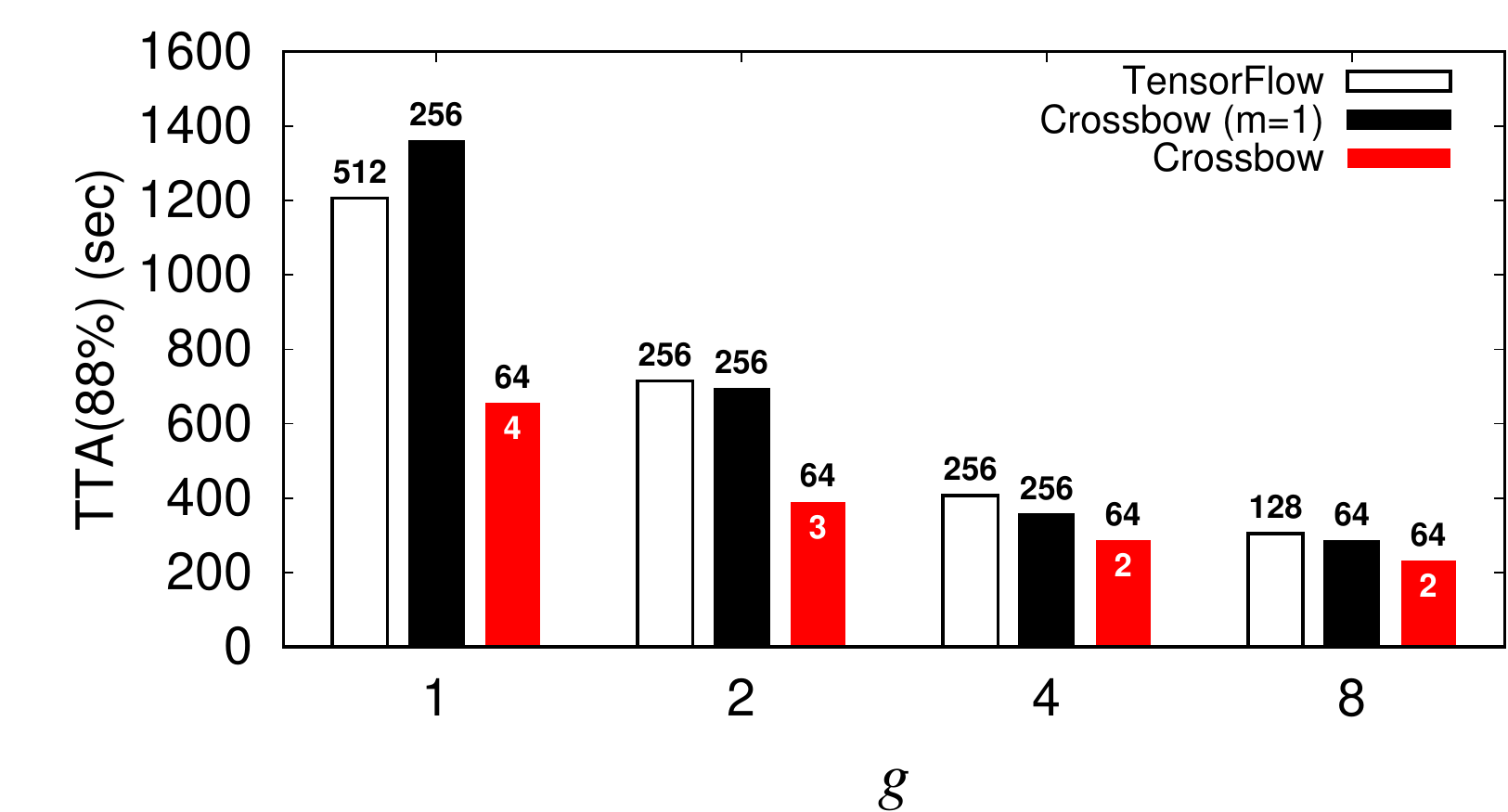}
\caption{ResNet-32}
\label{fig:scaling-resnet32}
\end{subfigure}
\begin{subfigure}[t]{0.325\textwidth}
  \includegraphics[width=\textwidth]{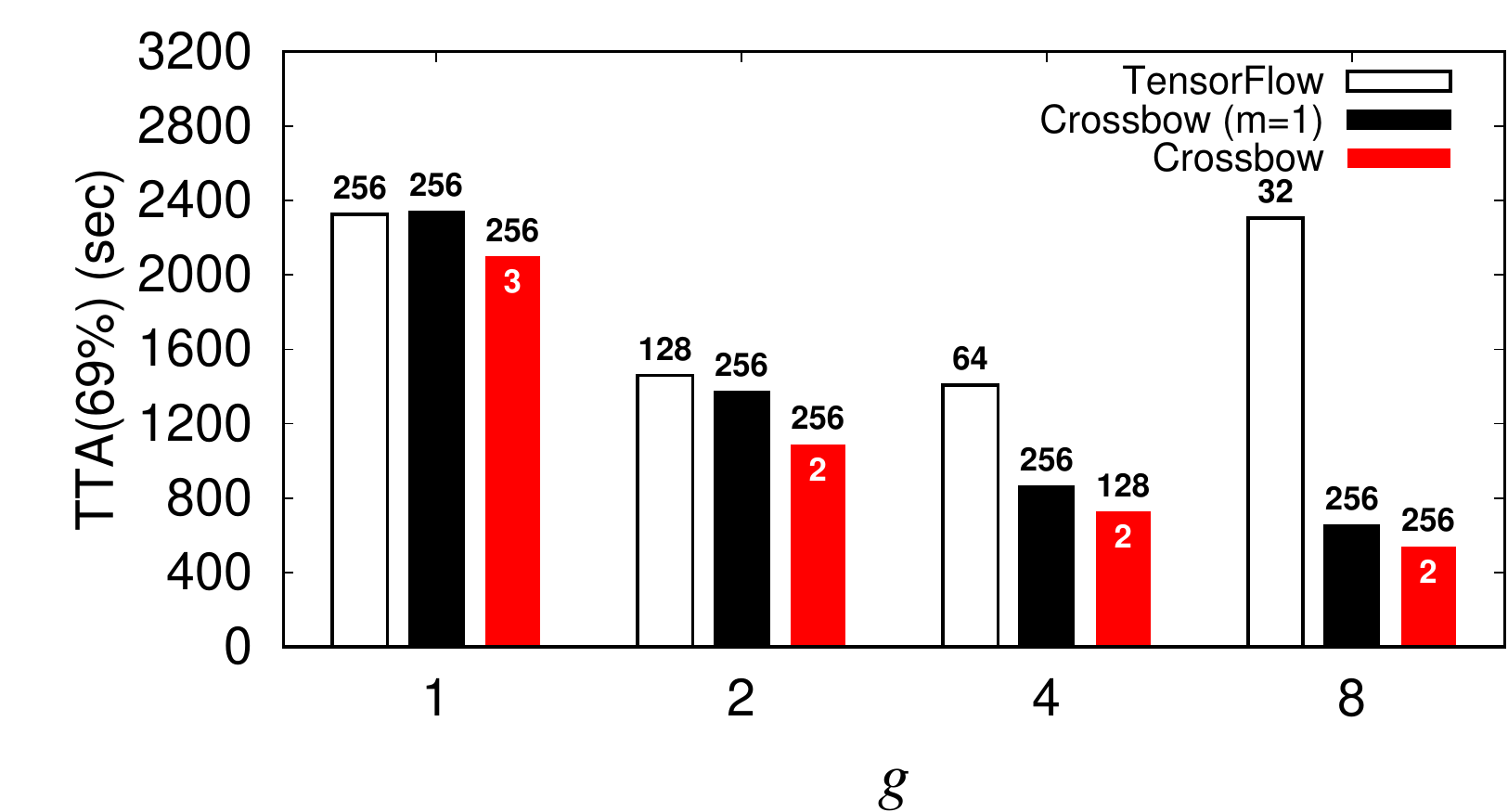}
  \caption{VGG}
  \label{fig:scaling-vgg}
\end{subfigure}
 \begin{subfigure}[t]{0.1625\textwidth}
    \includegraphics[width=\textwidth]{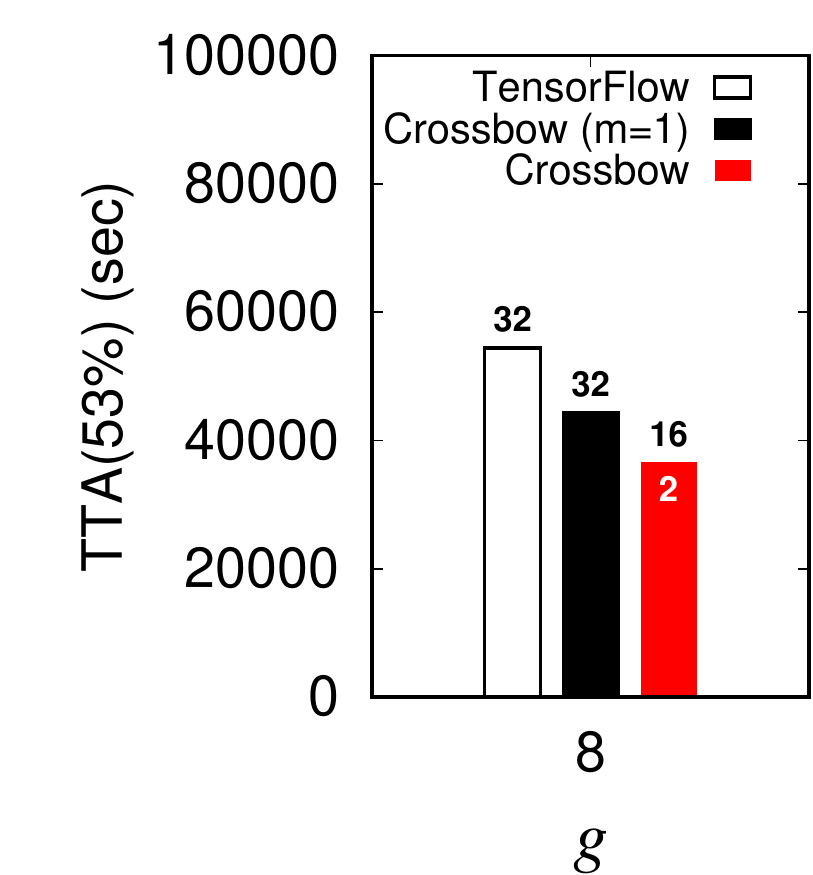}
    \caption{ResNet-50}
    \label{fig:scaling-resnet50}
  \end{subfigure} 
\begin{subfigure}[t]{0.1625\textwidth}
  \includegraphics[width=\textwidth]{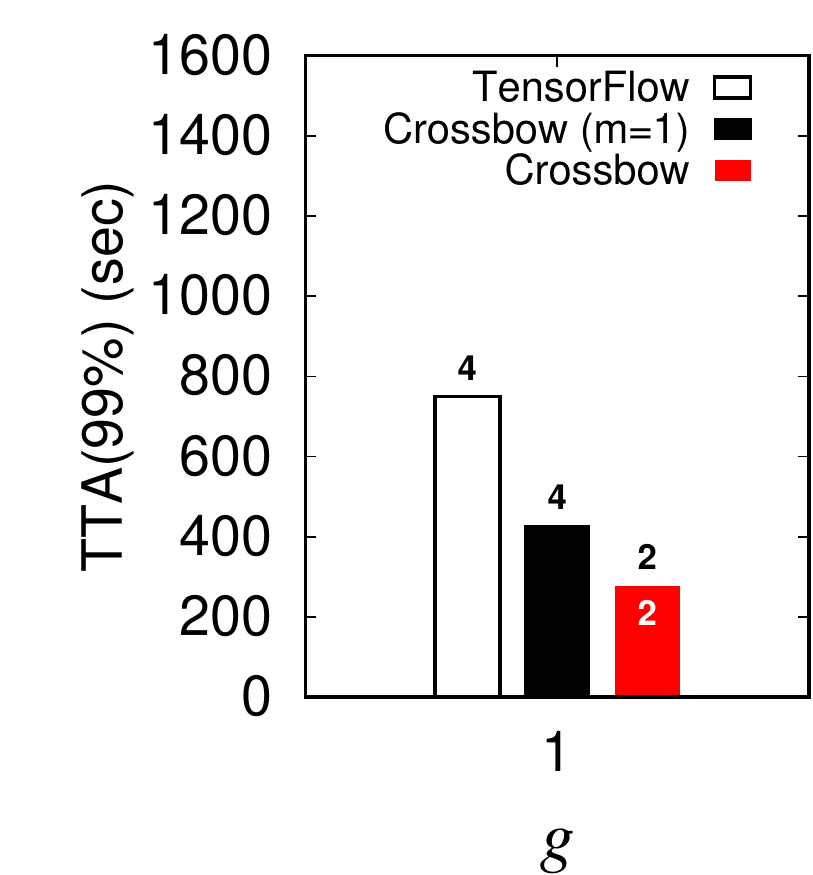}
  \caption{LeNet}
  \label{fig:scaling-lenet}
\end{subfigure}
\caption{Time-to-accuracy for four deep-learning models \captext{Numbers on top
    of the bars report the batch size per GPU that achieved that accuracy;
    numbers inside the \sys bars report the best number of model replicas per
    GPU.}}
\label{fig:scaling}
\end{figure*}

All experiments are conducted on a server with two~Intel Xeon~E5-2650~v3
2.3\unit{GHz} CPUs (20~CPU cores in total) and 256\unit{GB} of RAM. The server
has 8~GeForce GTX Titan~X (Pascal) GPUs, each with 3,072~cores and 12\unit{GB}
of RAM, connected via PCIe~3.0 ($\times$16). It runs Linux kernel~4.4 with the
NVIDIA driver~367.57 and CUDA~8.0 with the cuDNN~6.0 library.

As a baseline, we use TensorFlow version~1.4, and we compare its performance
against \sys using TensorFlow's suite of benchmarks~\cite{tf_benchmarks} shown
in Table~\ref{tab:apps}. We select a mix of models, representing different
sizes and shapes of networks. This includes both small~(LeNet) and large
networks~(ResNet-50) as well as deep and low-dimension networks~(ResNet-32 and
ResNet-50) and shallow and high-dimension ones~(VGG). To enable a fair
comparison, we configure both systems with the same data augmentation, model
variable initialisation and hyper-parameter settings.
Following common practices for training, in ResNet-32, the learning rate is
multiplied by 0.1 at epochs~80 and~120~\cite{HeZRS15}; in VGG, the learning
rate is halved every 20~epochs~\cite{geifmany2018}.

In our experiments, we vary two main parameters: the batch size per
learner~($b$) and the number of model replicas~($m$). Our main metric is the
time-to-accuracy \tta$(\mathsf{x})$, defined as the time taken for the median
test accuracy of the last 5~epochs to be equal or above a given
threshold~$\mathsf{x}$. For each of the four deep learning models, we choose
different thresholds based on the highest accuracy reached by TensorFlow in our
set-up. According to the results in~\Cref{fig:acc-target}, we select the
following thresholds: 99\%~(LeNet), 88\%~(ResNet-32), 69\%~(VGG-16) and
53\%~(ResNet-50). Higher accuracies can be achieved by leveraging dynamic
hyper-parameter tuning.
As discussed in~\S\ref{sec:problem-background}, these techniques, however, are
model- and architecture-specific and lack generality. In contrast, the goal of
our evaluation is to compare the different approaches underlying \sys and
TensorFlow under uniform settings.

\subsection{Scalability} %
\label{sec:eval:scaling}

\Cref{fig:scaling} shows the performance of \sys and TensorFlow when training
the four deep learning models. For ResNet-32 and VGG
(\Cref{fig:scaling-resnet32,fig:scaling-vgg}), we scale the number of GPUs from
1 to 8; for ResNet-50~(\Cref{fig:scaling-resnet50}), we show the results only
for 8~GPUs due to the long time required to train with fewer GPUs (\eg with
1~GPU, it would take more than 5~days with TensorFlow); for
LeNet~(\Cref{fig:scaling-lenet}), we only run the experiment with 1~GPU
because, given the small model size, training on multiple GPUs leads to an
increase in training time due to the synchronisation overhead.

First we consider the performance of \sys when using only one learner per GPU,
\ie
$m$$=$$1$ (black bar in~\Cref{fig:scaling}). In this configuration, for
ResNet-32 and VGG, \sys achieves a performance comparable or slightly worse
than TensorFlow when training on a small number of GPUs (1 or 2). The reason is
that both ResNet and VGG are relatively compute-intensive models. With few
learners, the synchronisation overhead is limited and, hence, the benefits of
\sync~(\S\ref{sec:sma}) and the fast task scheduling (\S\ref{sec:scheduling})
are less relevant. As we increase the number of GPUs to 4 or 8, the number of
learners and the amount of synchronisation among them increases
proportionally. The results now highlight the higher performance of \sync
compared to TensorFlow's S-SGD scheme, with up to a 72\% reduction in \tta for
VGG with 8~GPUs (and 7\% for ResNet-32, respectively). A similar improvement
can be observed for ResNet-50 in~\Cref{fig:scaling-resnet50}, with \sys
achieving a 18\% reduction in \tta with 8~GPUs.

\sys, however, can offer a benefit even when training on a single GPU (one
learner in total) if the model is not compute-intensive. For LeNet, each
learning task takes 1\unit{ms} or less (compared to $\sim$220\unit{ms} for
ResNet-50) and, hence, the scheduling overhead becomes more critical. By
leveraging its efficient task scheduler, \sys yields a significant \tta
reduction (43\%) compared to TensorFlow with one
learner~(see~\Cref{fig:scaling-lenet}).

Thus far, we have focused on a \sys configuration with
$m$$=$$1$. A key advantage of \sys is its ability to increase hardware
efficiency without affecting statistical efficiency by adding more learners per
GPU
($m$$>$$1$). In this case, \sys significantly improves performance even when
training with few GPUs. For ResNet-32, \sys with
$m$$=$$4$ achieves a 46\% \tta reduction with 1~GPU and a 24\% \tta reduction
with
$m$$=$$2$ and 8~GPUs; for VGG, the reduction is 10\% for 1~GPU and 77\% for
8~GPUs. Similar benefits with
$m$$=$$2$ also exist for ResNet-50 (33\% \tta reduction, corresponding to
5\unit{hours}) and LeNet (63\% \tta reduction). As explained in
\S\ref{sec:sma}, the reason is that
$m$$>$$1$ increases the GPU throughput without necessitating a larger batch
size that would affect statistical efficiency. For ResNet-50, \sys with
$m$$=$$2$ uses a small batch size of
$b$$=$$16$ compared to an aggregate batch size of
$32$$\times$$8$$=$$256$ for TensorFlow (64~and 1,024~for ResNet-32,
respectively).

\begin{figure}
\centering
\begin{subfigure}[t]{0.49\columnwidth}
\includegraphics[width=\textwidth]{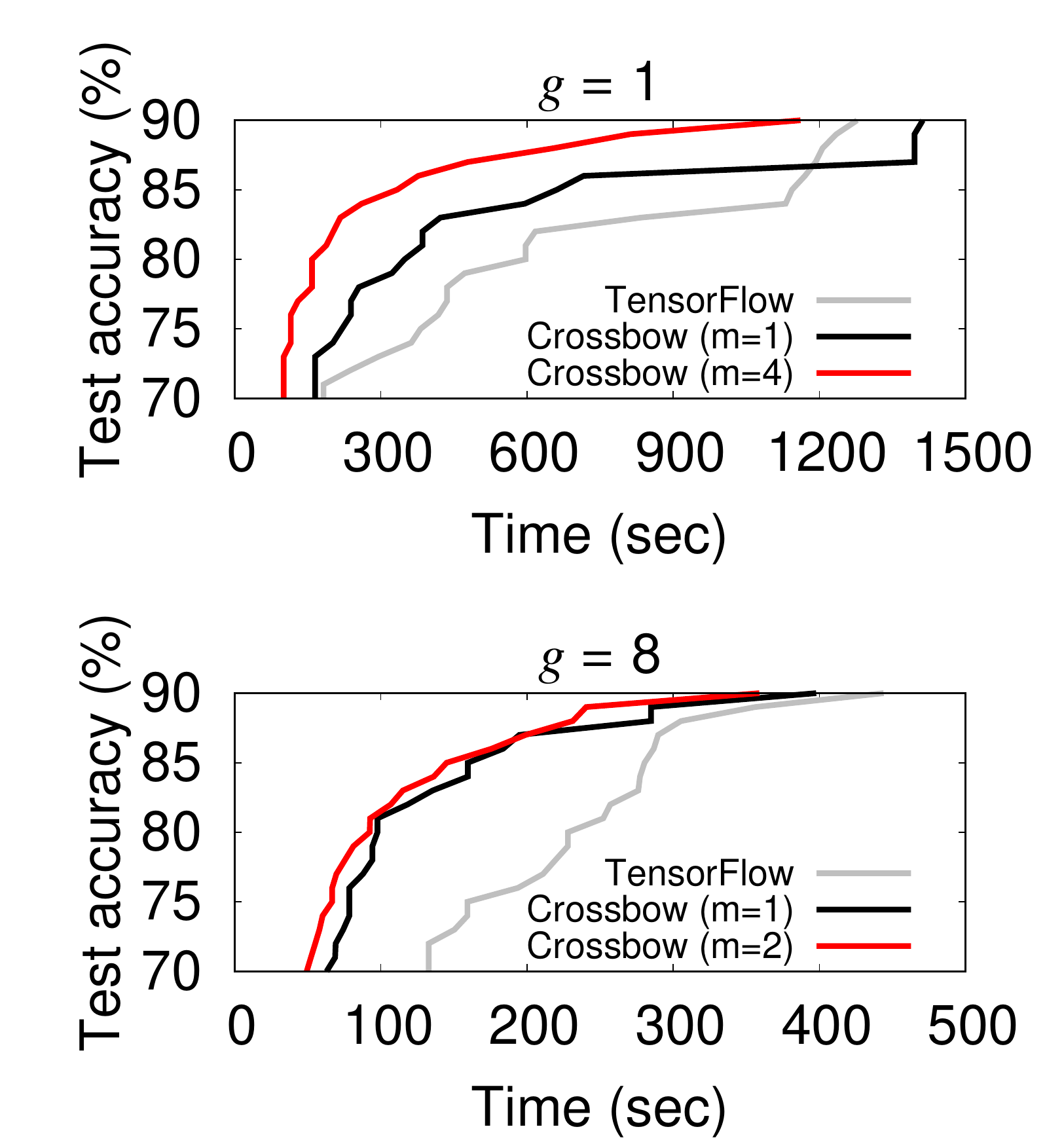}
\caption{ResNet-32}
\label{fig:acc-over-time-resnet}
\end{subfigure}
\begin{subfigure}[t]{0.49\columnwidth}
\includegraphics[width=\textwidth]{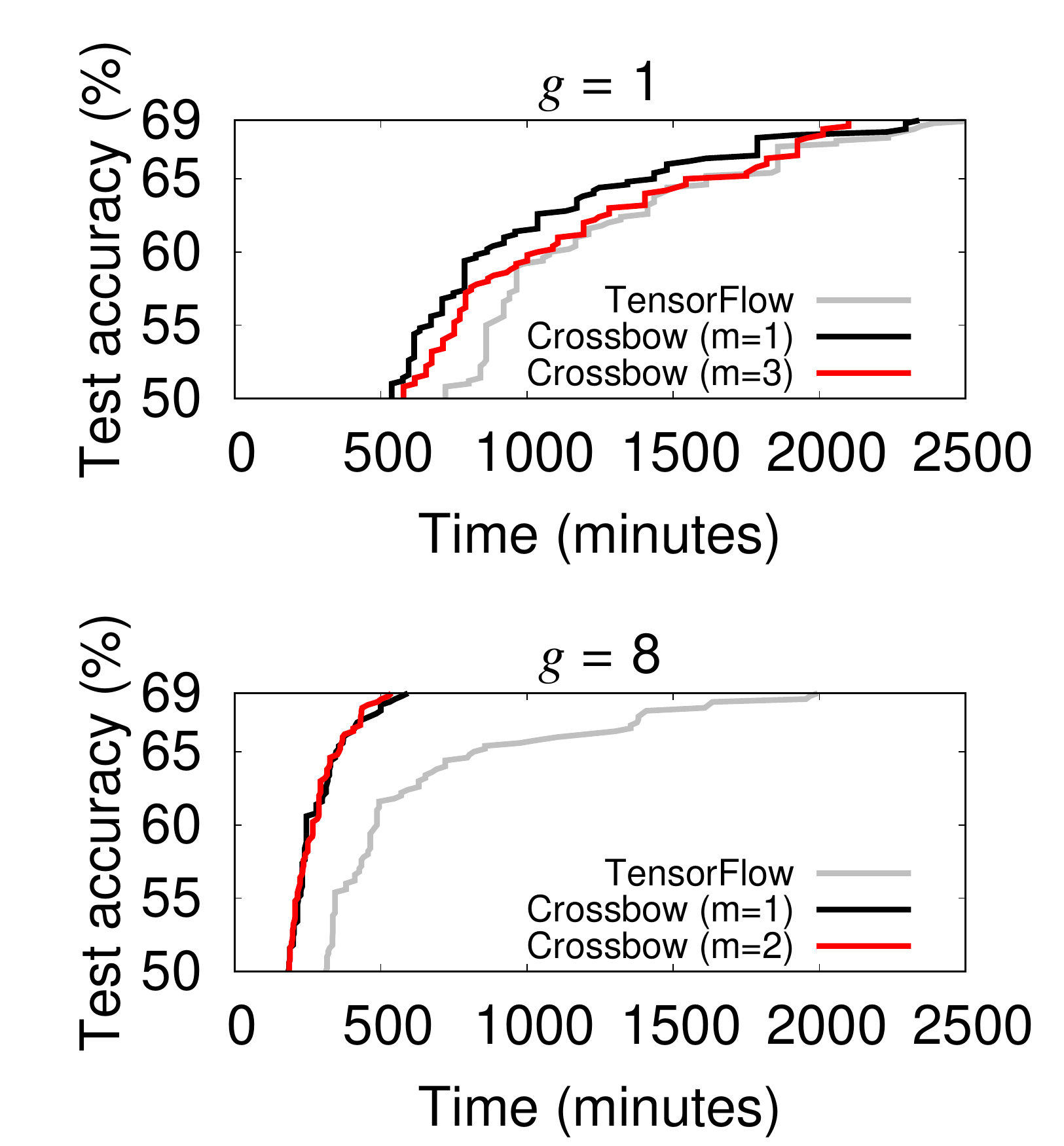}
\caption{VGG}
\label{fig:acc-over-time-vgg}
\end{subfigure}
\caption{\mbox{Convergence with 1~(top) and 8 (bottom)~GPUs}}\label{fig:acc-over-time}
\end{figure}

While we choose to use the highest accuracy reached by TensorFlow as the
threshold~$\mathsf{x}$ in the \tta$\mathsf{(x)}$ metric, similar improvements
hold for other accuracy thresholds. In~\Cref{fig:acc-over-time}, we plot \tta
over time for both ResNet-32 and VGG with 1~and 8~GPUs, respectively. \sys
achieves high accuracy within a few minutes: with 8~GPUs, \sys takes
92\unit{seconds} to exceed a 80\% accuracy for ResNet-32 compared to
252\unit{seconds} for TensorFlow---a 63\% \tta reduction. Similarly, \sys
achieves a 74\% \tta reduction for VGG. This indicates that \sync converges
quickly to a region containing good minima.

\subsection{Statistical and hardware efficiency}
\label{sec:eval:batch-size}

\begin{figure*}[tb]
  \centering  
  \begin{subfigure}[!t]{0.325\textwidth}
    \includegraphics[width=\textwidth]{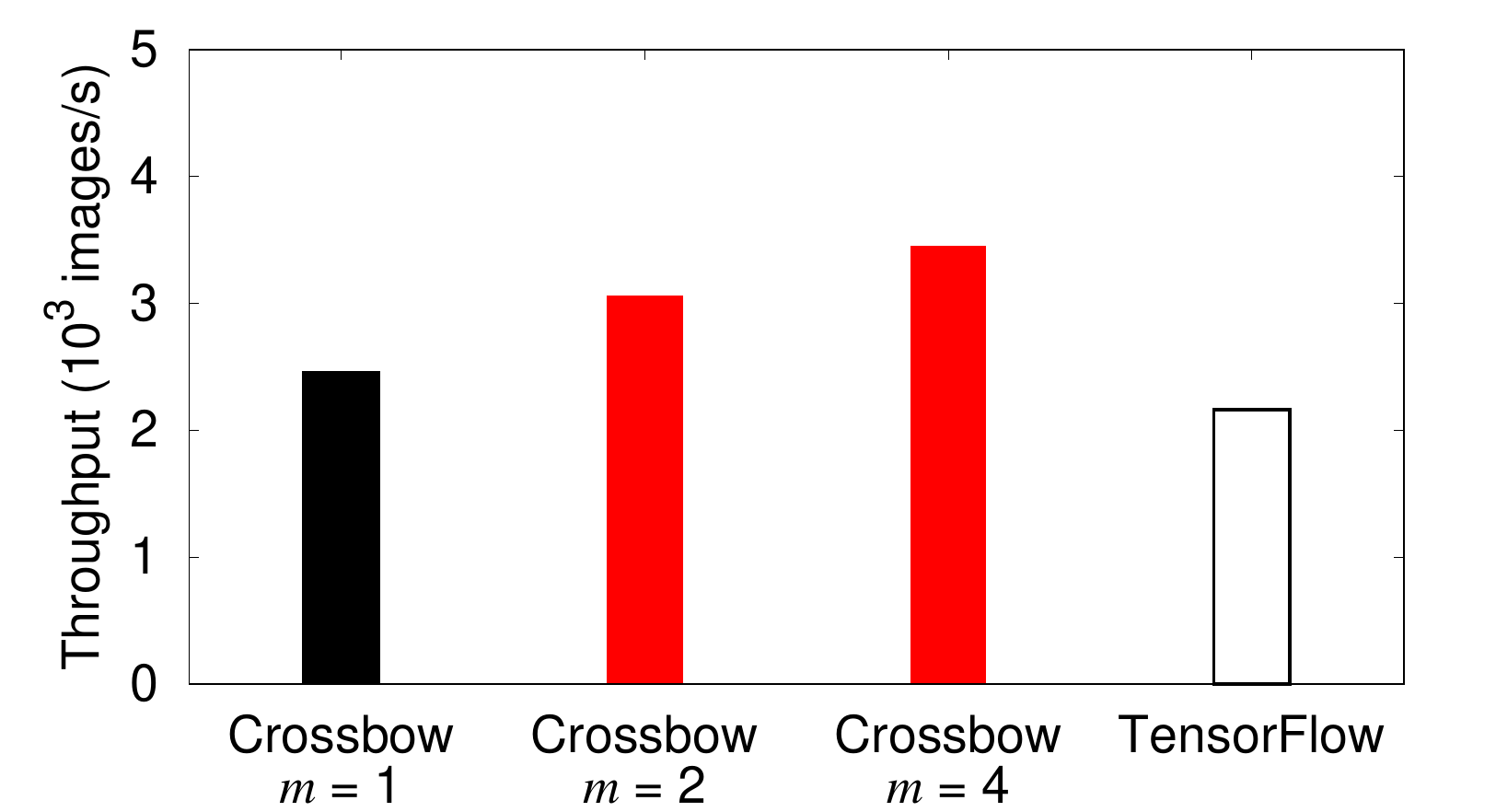}
    \caption{Hardware efficiency}
    \label{fig:hw-efficiency-1}
  \end{subfigure}
  \begin{subfigure}[!t]{0.325\textwidth}
    \includegraphics[width=\textwidth]{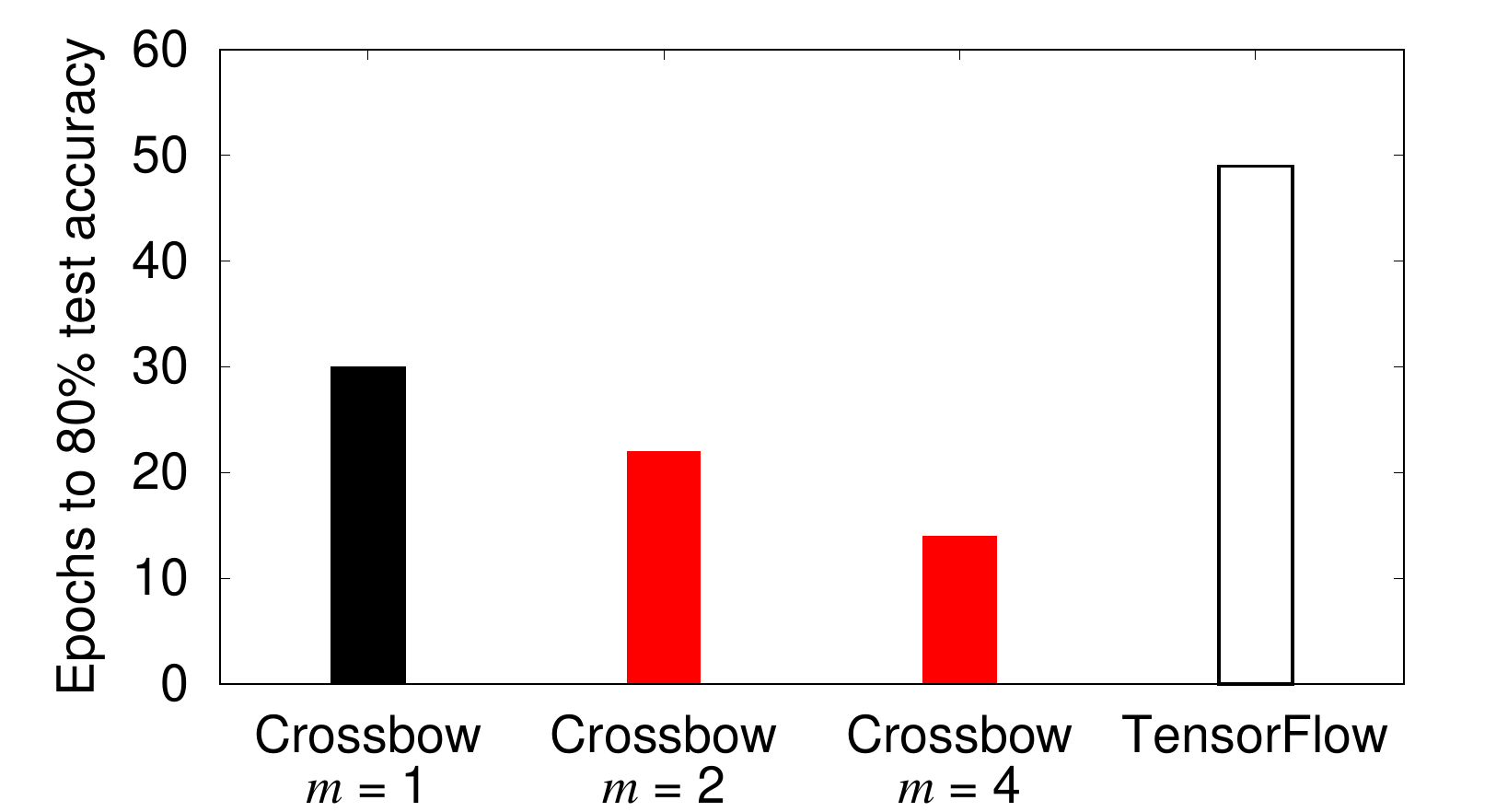}
    \caption{Statistical efficiency}
    \label{fig:stat-efficiency-1}
  \end{subfigure}
  \begin{subfigure}[!t]{0.325\textwidth}
    \includegraphics[width=\textwidth]{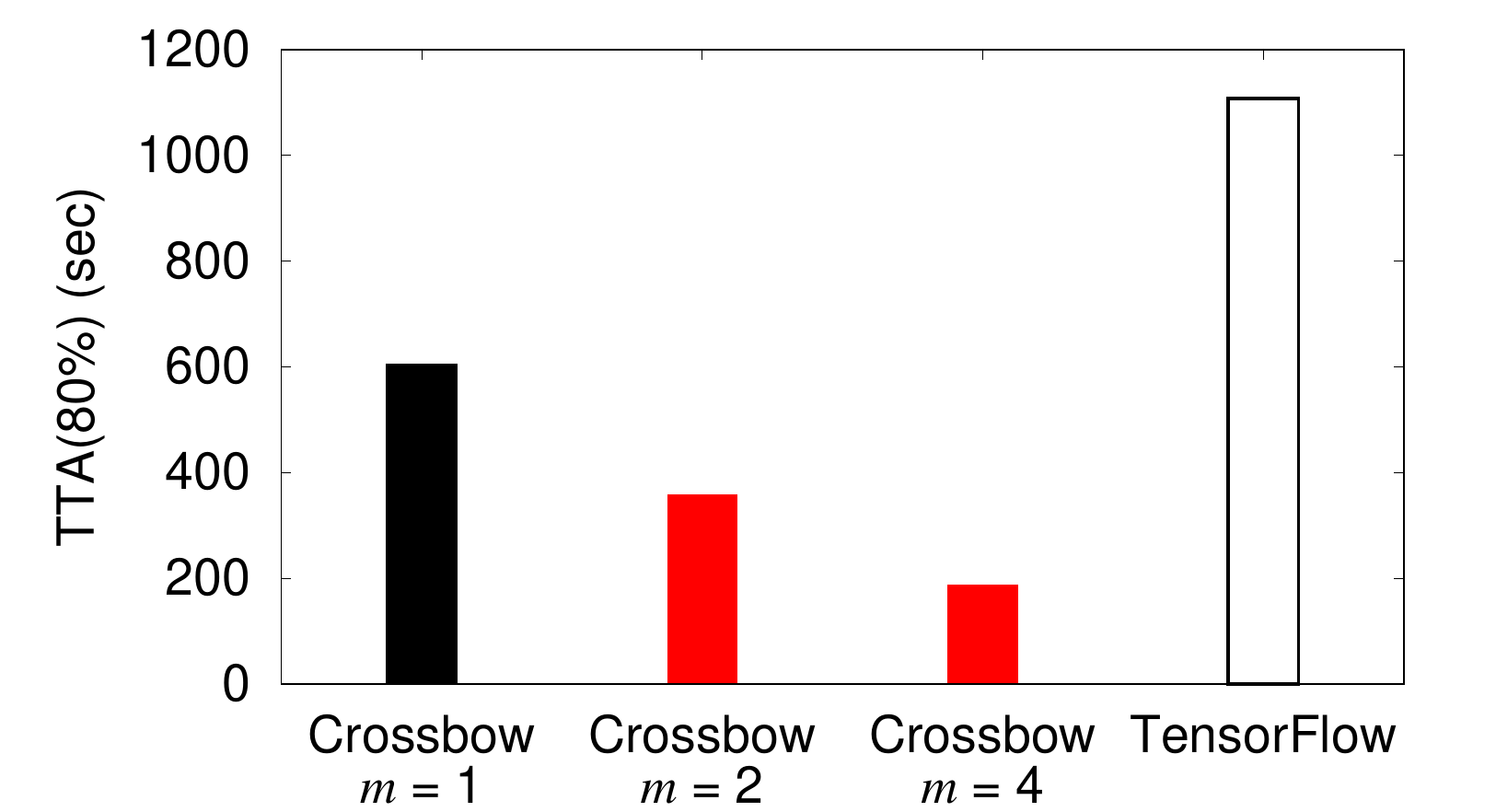}
    \caption{Time to accuracy}
    \label{fig:stat-hw-efficiency_tta-1}
  \end{subfigure}
  \caption{Trade-off between hardware and statistical efficiency with 1 GPU
    \captext{This experiment uses ResNet-32 and $b$$=$$64$.}}
  \label{fig:stat-hw-efficiency-1}
\end{figure*}

\begin{figure*}[tb]
  \centering  
  \begin{subfigure}[!t]{0.325\textwidth}
    \includegraphics[width=\textwidth]{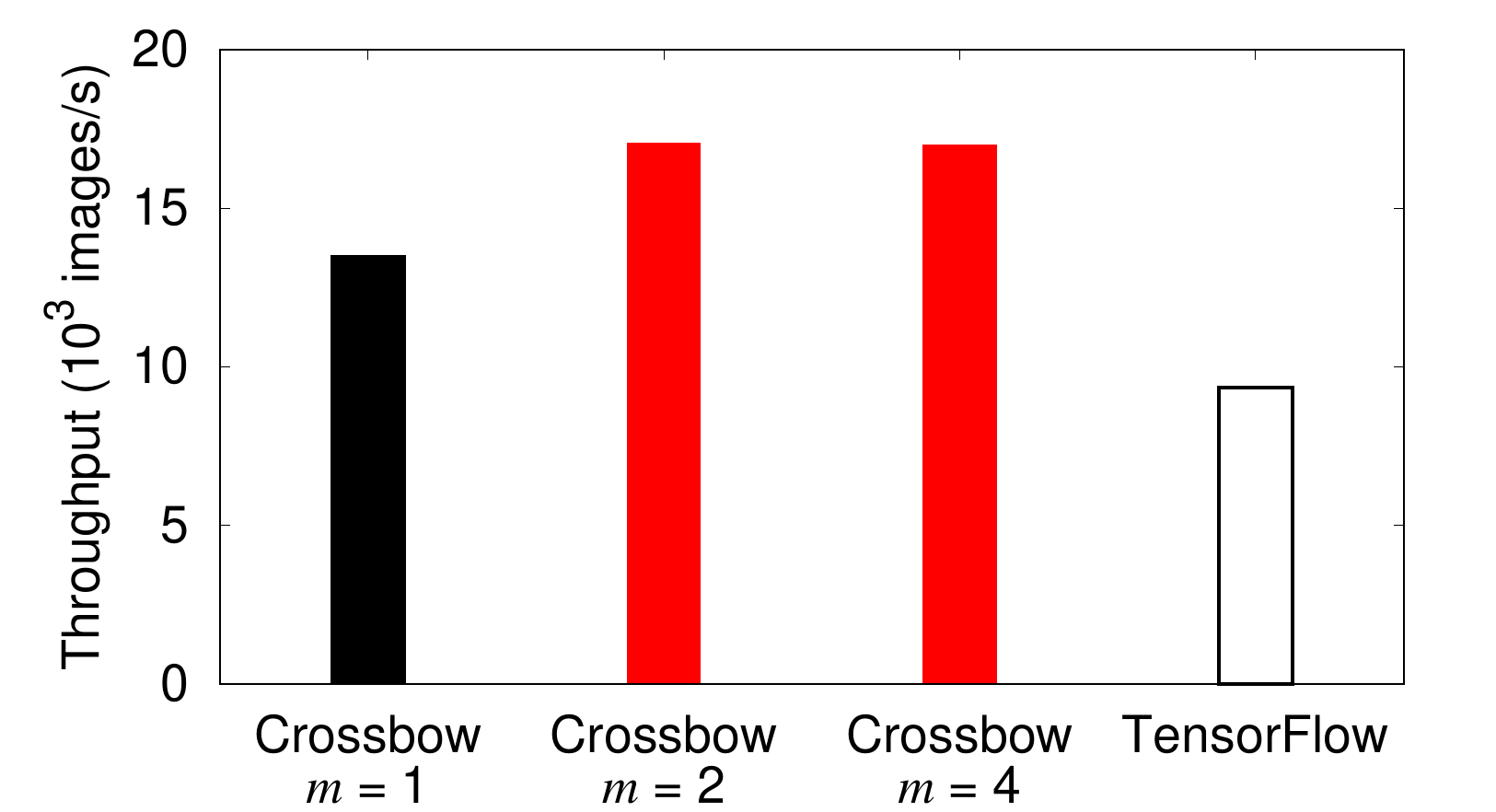}
    \caption{Hardware efficiency}
    \label{fig:hw-efficiency-8}
  \end{subfigure}
  \begin{subfigure}[!t]{0.325\textwidth}
    \includegraphics[width=\textwidth]{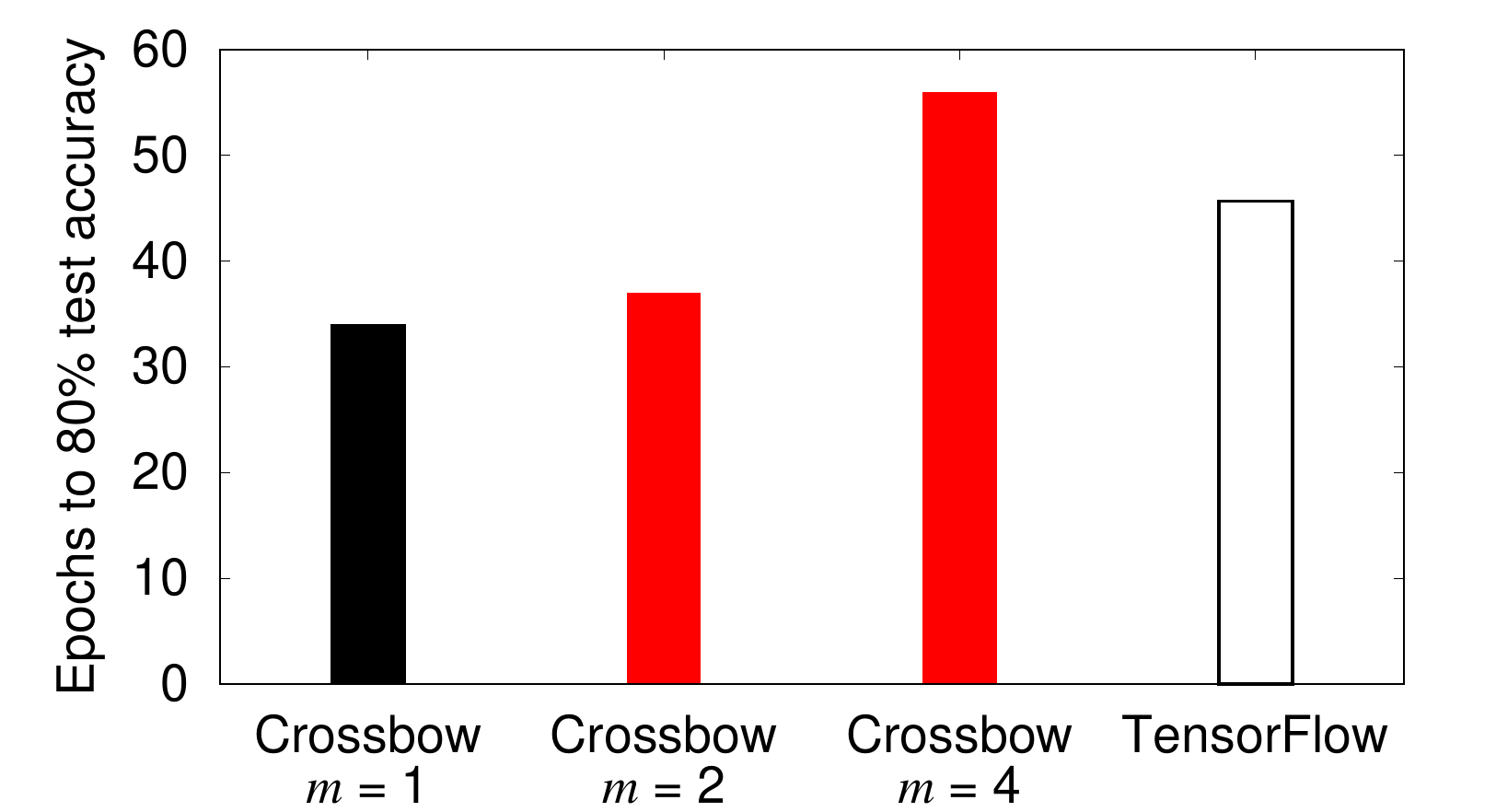}
    \caption{Statistical efficiency}
    \label{fig:stat-efficiency-8}
  \end{subfigure}
  \begin{subfigure}[!t]{0.325\textwidth}
    \includegraphics[width=\textwidth]{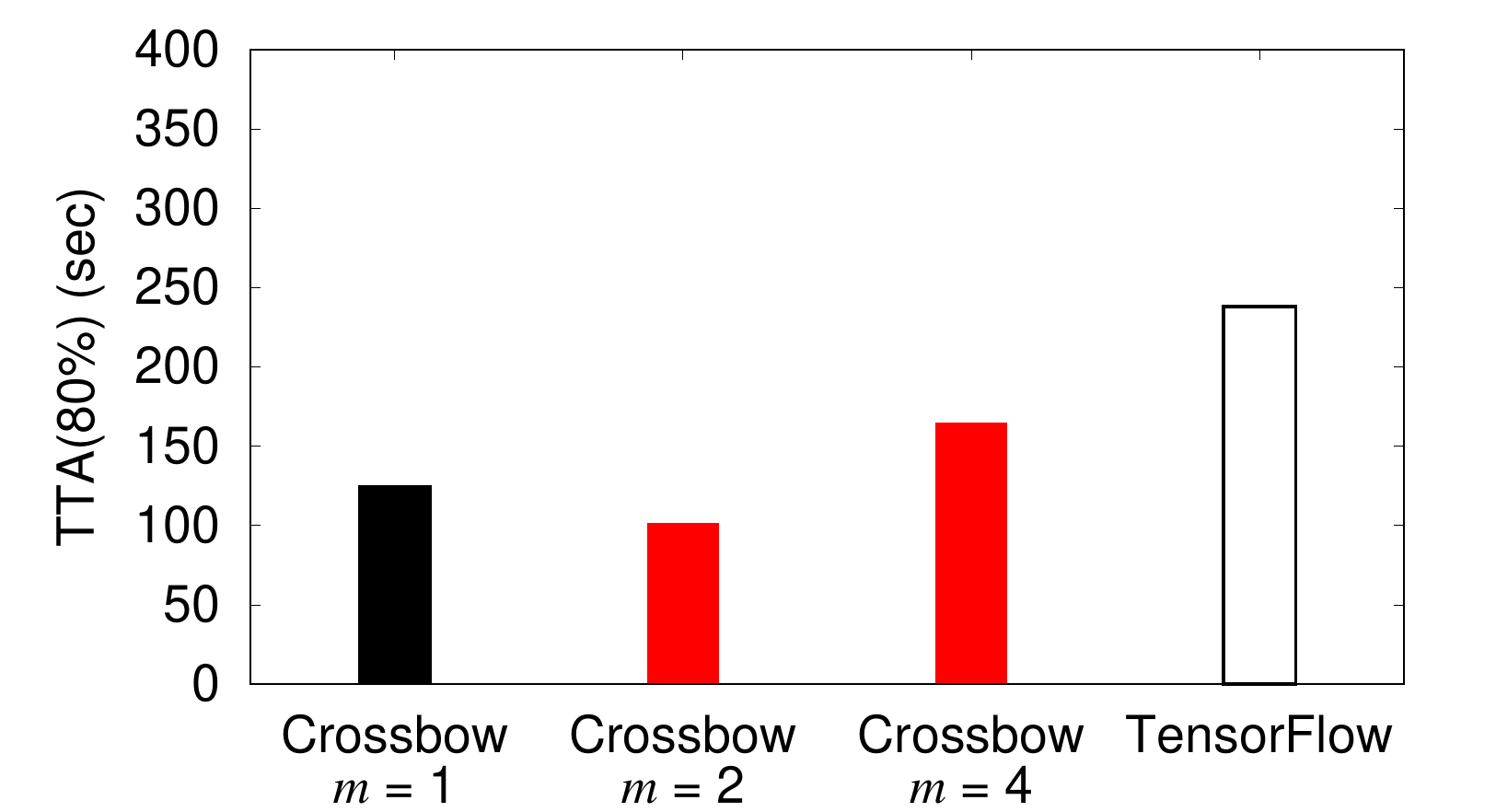}
    \caption{Time to accuracy}
    \label{fig:stat-hw-efficiency_tta-8}
  \end{subfigure}

  \caption{Trade-off between hardware and statistical efficiency with 8 GPUs
    \captext{This experiment uses ResNet-32 and $b$$=$$64$.}}
  \label{fig:stat-hw-efficiency-8}
\end{figure*}

\sys with
$m$$>$$1$ outperforms TensorFlow because it can increase hardware efficiency
without negatively impacting statistical efficiency. As discussed
in~\S\ref{sec:problem-background}, the only way to improve hardware efficiency
in TensorFlow is to increase the batch size but this comes at the cost of
reduced statistical efficiency. In contrast, \sys can use the number of
learners~$m$ as an additional control parameter to increase hardware efficiency
without having to resort to larger batch sizes. We investigate this in more
detail in~\Cref{fig:stat-hw-efficiency-1,fig:stat-hw-efficiency-8}, in which we
show how hardware and statistical efficiency, and the resulting \tta, are
affected by $m$ when using 1~and 8~GPUs respectively. We only report the
results for
ResNet-32~($b$$=$$64$) but similar trends hold for the other models. Compared
to the experiments in~\Cref{fig:scaling-resnet32}, we lower the target accuracy
for \tta to 80\% as otherwise the results would be skewed by the change in the
learning rate at epoch 80 (see~\S~\ref{sec:eval:settings}).

When training with 1~GPU, using
$m$$=$$4$ increases the throughput by a factor of 1.4$\times$ compared to the
case with a single learner~(\Cref{fig:hw-efficiency-1}) because the multiple
learners fully utilise a GPU. Interestingly, this improves statistical
efficiency as well---the number of epochs required to converge drop from
30~($m$$=$$1$) to
14~($m$$=$$4$) (see~\Cref{fig:stat-efficiency-1}). This is because multiple
model replicas can explore a larger portion of the space in parallel while the
average model can reduce the variance among them, thus requiring fewer epochs
to find good minima. As a result of the higher hardware \emph{and} statistical
efficiencies, the \tta is also reduced by
3.2$\times$~(\Cref{fig:stat-hw-efficiency_tta-1}).

In contrast, the behaviour with 8~GPUs is somewhat different. While
$m$$=$$2$ yields higher throughput (1.3$\times$), increasing the number of
learners to
$m$$=$$4$ does not further improve the
throughput~(\Cref{fig:hw-efficiency-8}). The reason is that, with 8~GPUs and
4~learners per GPU, the overall number of learners is 32, which introduces a
significant amount of synchronisation overhead. In terms of statistical
efficiency~(\Cref{fig:stat-efficiency-8}), increasing the number of learners to
$m$$=$$2$ does not significantly affect the number of epochs to converge but
increasing it further leads to reduced statistical efficiency---with
32~learners in total, there is not enough stochastic noise in the training
process, which makes it harder for the average model to escape bad minima. In
this case,
$m$$=$$2$ represents the best trade-off because it allows for higher hardware
efficiency without noticeably worsening statistical efficiency. Indeed, this
configuration reduces training time by
1.3$\times$~(\Cref{fig:stat-hw-efficiency_tta-8}).

These results show that increasing the number of learners per GPU is beneficial
to reduce the training time. However, identifying the correct number of
learners is crucial to achieving the best performance, as we show next.

\subsection{Selecting the number of learners}
\label{sec:eval:auto-tuning}

\begin{figure}[!t]
\centering
\begin{subfigure}[t]{0.49\columnwidth}
\includegraphics[width=\textwidth]{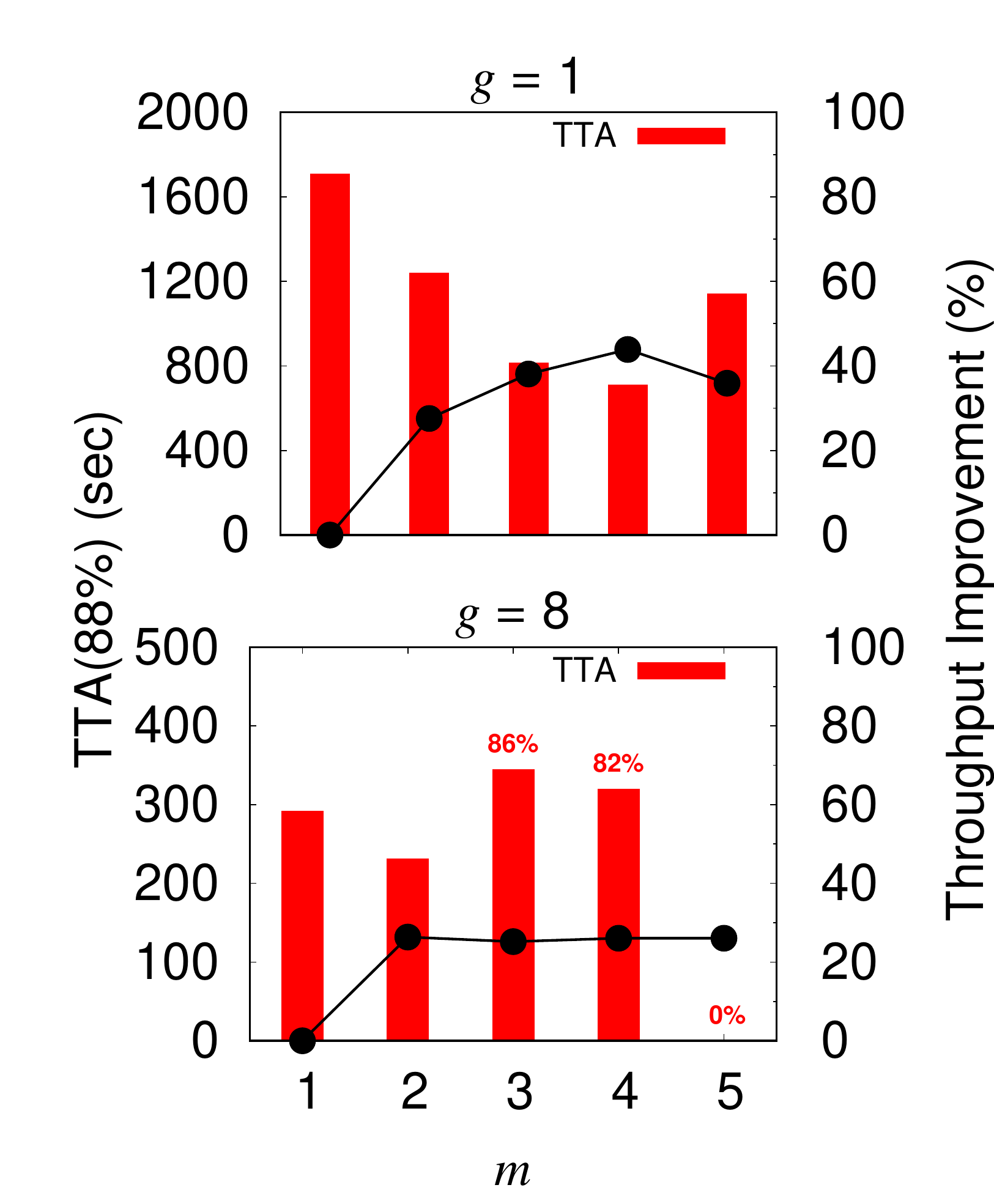}
\caption{ResNet-32, $b = 64$}
\label{fig:auto-tuning-resnet32}
\end{subfigure}
\begin{subfigure}[t]{0.49\columnwidth}
\includegraphics[width=\textwidth]{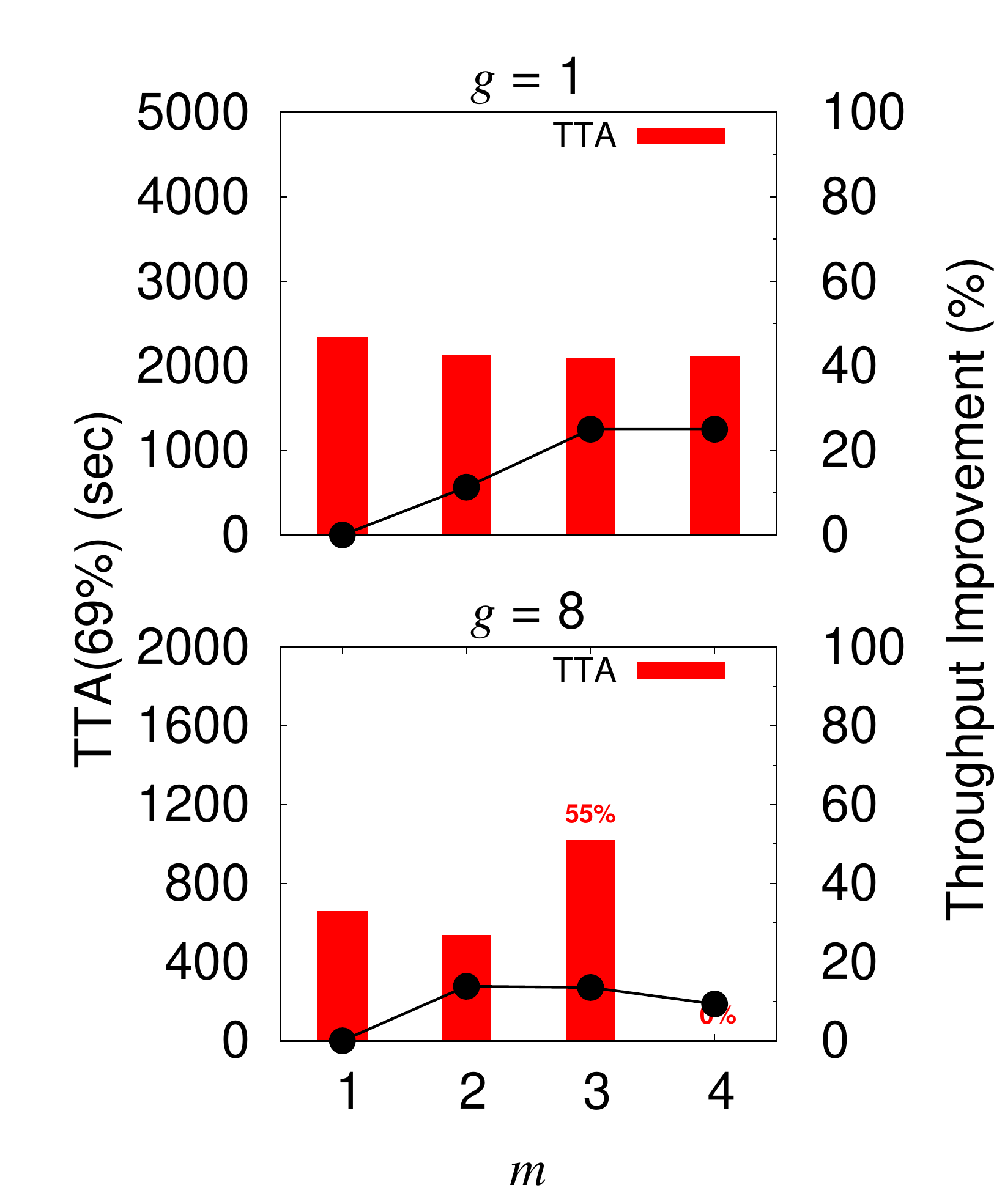}
\caption{VGG, $b=256$}
\label{fig:auto-tuning-vgg}
\end{subfigure}
\caption{Varying the number of models \captext{This experiment uses the same
    batch sizes as~\Cref{fig:scaling}. The best number of models (\ie
    minimising \tta) is the one that saturates training throughput.}}
\label{fig:auto-tuning}
\end{figure}

To select the number of learners per GPU~$m$, \sys progressively increases $m$
until the throughput (expressed as images processed per second) stops improving
(see~\Cref{sec:autotuning}). To validate this approach, \Cref{fig:auto-tuning}
shows the \tta and throughput achieved for increasing values of $m$ when
training ResNet-32 and VGG with 1~and 8~GPUs, respectively.

For ResNet-32, we observe that, with 1~GPU, the throughput grows until
$m$$=$$4$ and then decreases; with 8~GPU, the throughput improves until
$m$$=$$2$ and then remains relatively constant for
$m$$>$$2$. As predicted by our auto-tuning technique, the lowest \tta is
achieved with $m$$=$$4$ for 1~GPU and
$m$$=$$2$ for 8~GPUs, respectively. Similarly, for VGG, the throughput plateaus
at $m$$=$$3$ for 1~GPU and
$m$$=$$2$ for 8~GPUs, respectively, which correspond to the lowest values of
\tta. This demonstrates the ability of \sys{}'s auto-tuning technique to
identify quickly the correct number of learners in order to minimise \tta.

\subsection{Synchronisation model}
\label{sec:eval:sma}

\begin{figure}[t]
\centering
\includegraphics[width=0.9\columnwidth]{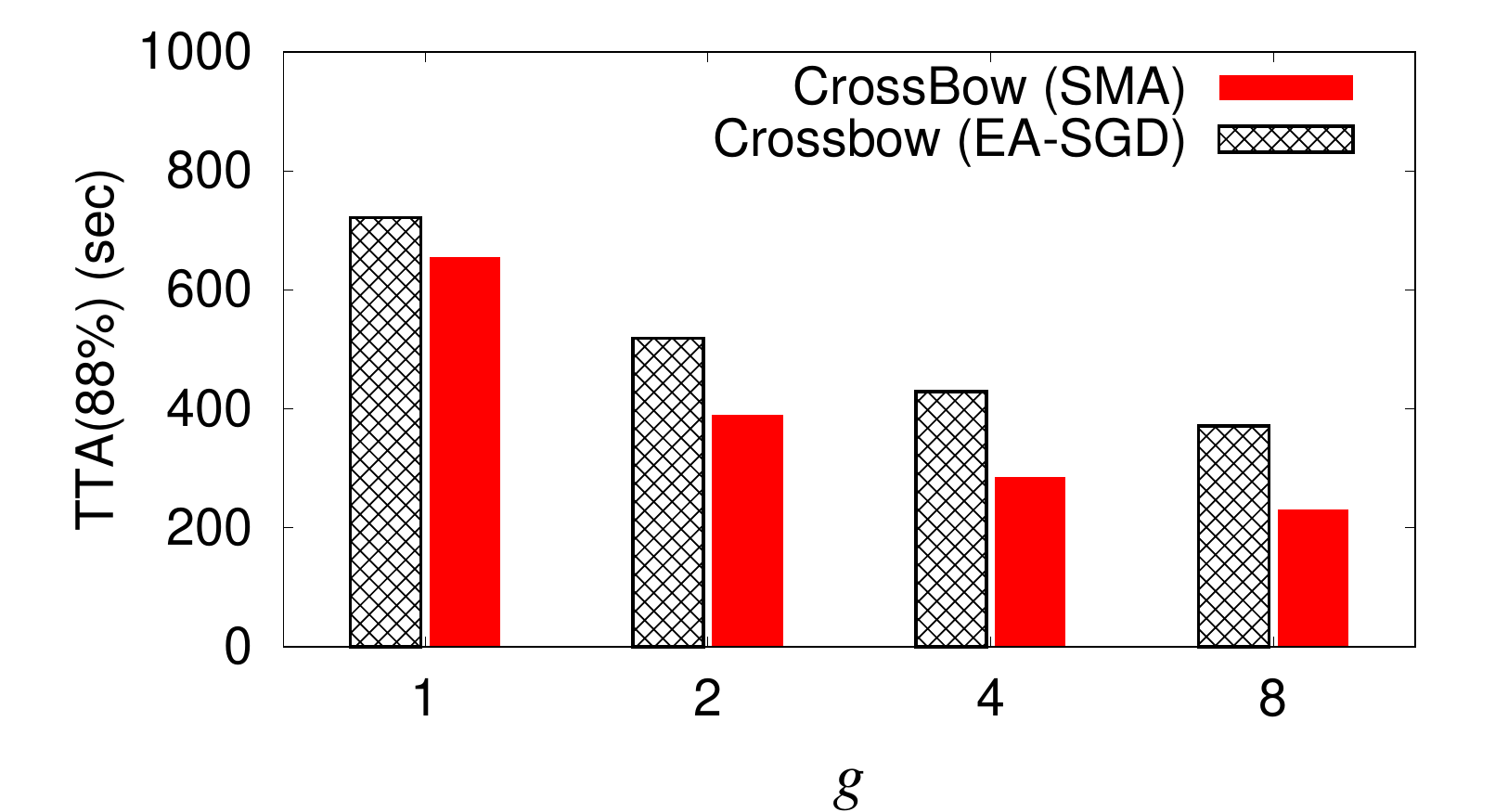}
\caption{Synchronisation with \sync compared to EA-SGD
  in \sys \captext{This experiment uses ResNet-32.}}
\label{fig:sma-compare}
\end{figure}

\begin{figure}[t]
\centering
\includegraphics[width=0.9\columnwidth]{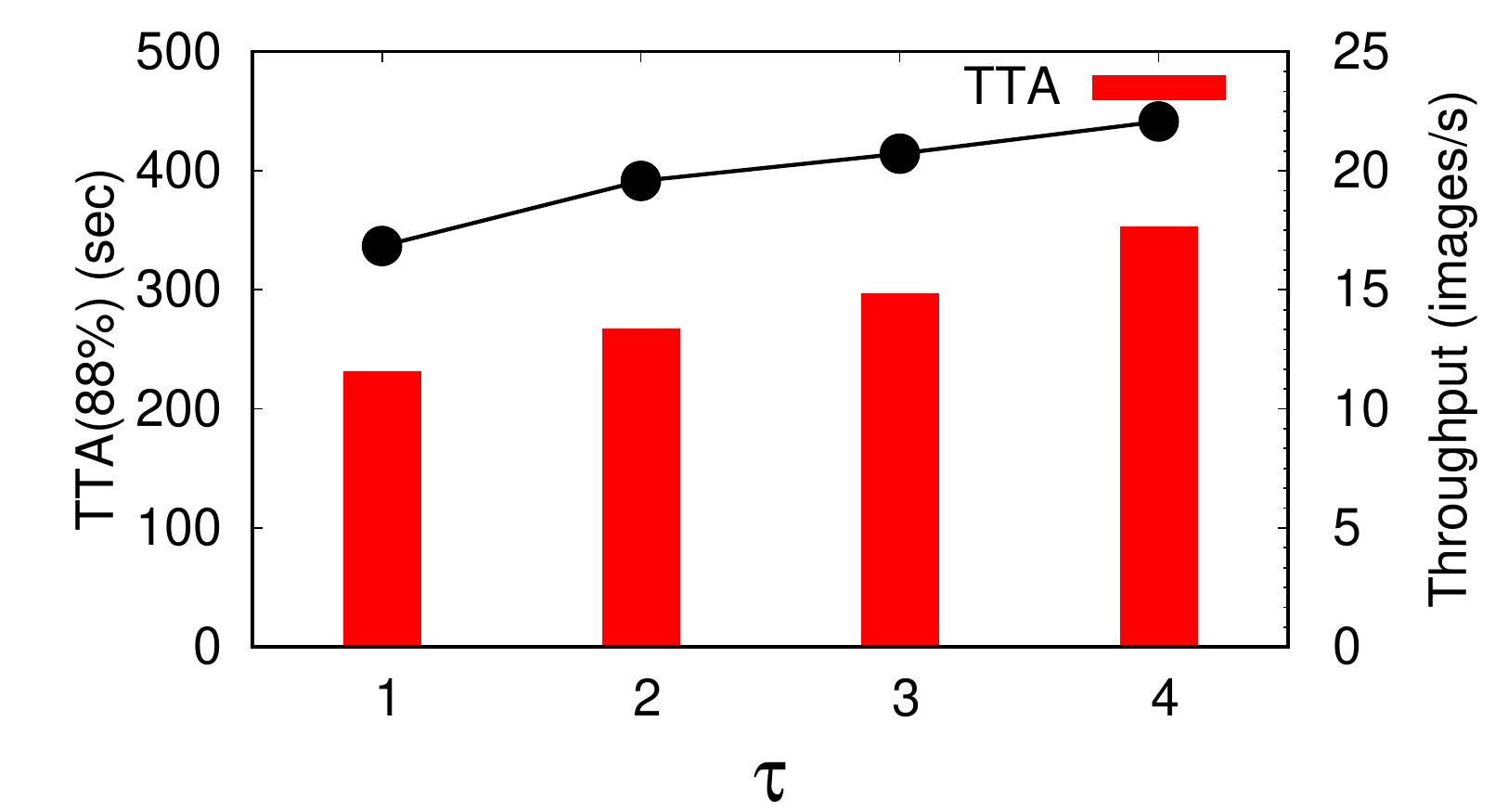}
\caption{Effect of synchronisation frequency on time-to-accuracy in \sys
  \captext{This experiment uses ResNet-32, $\gpuvar$$=$$8$ and $m$$=$$2$.}}
\label{fig:sma-frequency}
\end{figure}

Now we turn our attention to \sync, the synchronisation approach used by
\sys{}, by comparing it to elastic averaging SGD~(\easgd)~\cite{Zhang2015}.
To quantify the impact of momentum in \sync, we compare the performance of \sys
when training ResNet-32 using \SMA against the performance using \easgd for an
increasing number of GPUs.

\Cref{fig:sma-compare} shows that \SMA reduces \tta compared to \easgd by 9\%
with 1~GPU and $m$$=$$4$ and by 61\% with 8~GPUs and
$m$$=$$2$ (16~learners). The reason why the gap increases with the number of
GPUs is that the more learners are used, the lower the (asymptotic) variance of
the average model becomes, which makes it hard to escape from local
minima. Therefore, without including momentum, the average model converges more
slowly.

\sys synchronises the different replicas with the average model in each
iteration. The authors of \easgd propose to synchronise every
$\uptau$$>$$1$ iterations to reduce the communication overhead. We study the
impact of this optimisation on \sys in~\Cref{fig:sma-frequency}. While less
frequent synchronisation
($\uptau$$>$$1$) increases the overall throughput (up to 31\% for
$\uptau$$=$$4$ compared to
$\uptau$$=$$1$), it negatively affects convergence speed, resulting in a higher
\tta (53\% longer with $\uptau$$=$$4$ compared to
$\uptau$$=$$1$). Therefore, we always use $\uptau$$=$$1$.

\subsection{Synchronisation efficiency}
\label{sec:eval:overhead}

\begin{figure}[t]
\centering
\includegraphics[width=0.9\columnwidth]{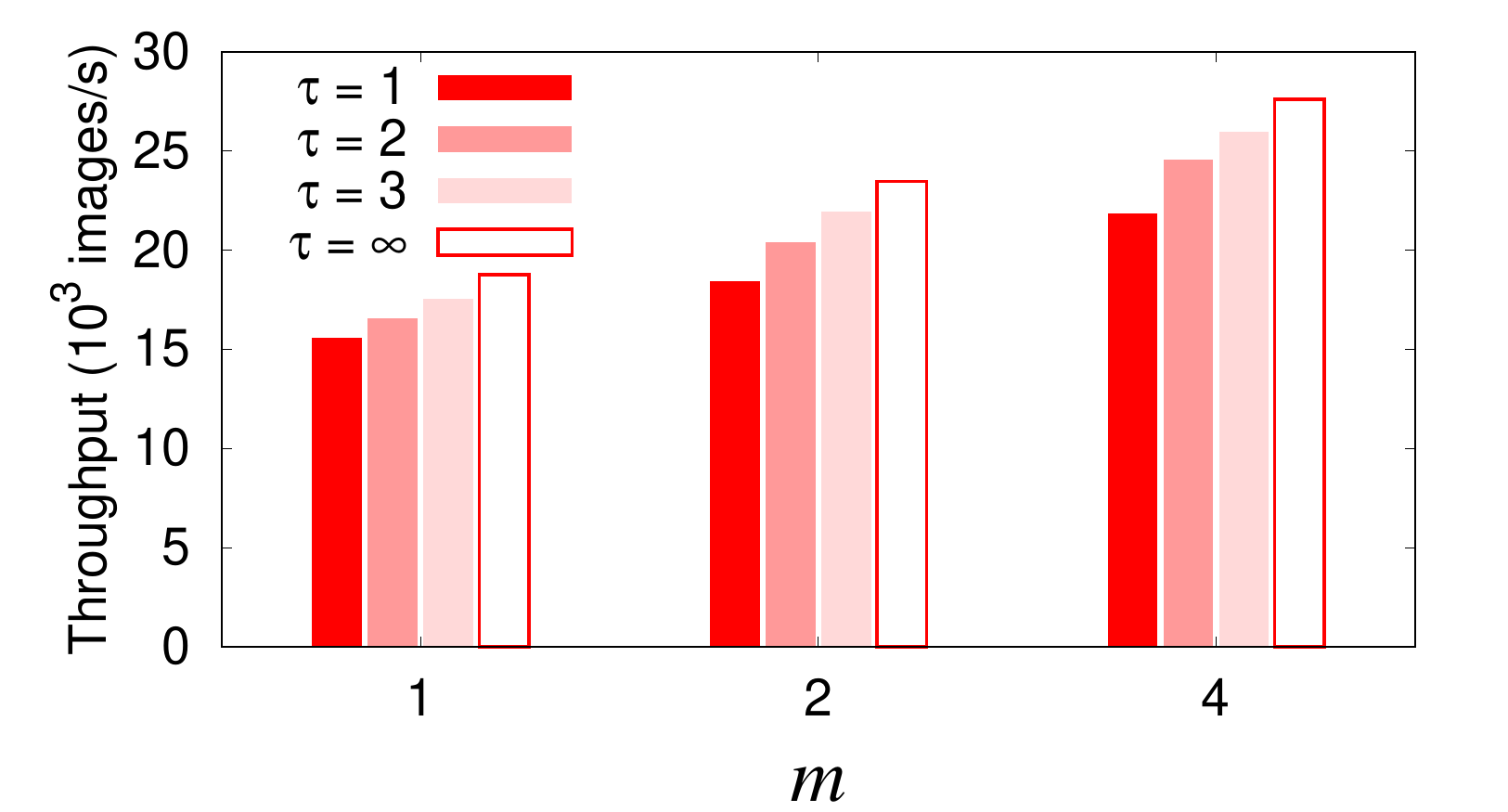}
\caption{Effect of synchronisation frequency on hardware efficiency in \sys 
  \captext{This experiment uses ResNet-32 and $\gpuvar$$=$$8$.}}
\label{fig:overhead}
\end{figure}

In the previous set of experiments, we focused on the \emph{algorithmic}
benefits of the \sync synchronisation approach. Now, instead, we want to
understand the performance of our synchronisation
\emph{implementation}. Similar to~\Cref{fig:sma-frequency}, we conduct an
experiment in which we measure the throughput achieved by \sys for increasing
values of $\uptau$, including also the case with \emph{no} synchronisation at
all. Since we are only interested in the performance aspect, we do not report
the \tta (we have already shown that
$\uptau$$=$$1$ yields the shortest \tta). The goal of this experiment is rather
to observe the increase in throughput as we reduce the synchronisation
frequency. The rationale is that, if synchronisation incurred a high cost, the
throughput would drastically increase as we reduce the amount of
synchronisation.

Contrary to this expectation, the results in~\Cref{fig:overhead} show that
throughput is only marginally higher, growing from 15,500\unit{images/s} with
$\uptau$$=$$1$ to 18,500\unit{images/s} with no synchronisation at all (20\%)
with $m$$=$$1$ (27\% for $m$$=$$4$, respectively). This indicates that \sys{}'s
synchronisation implementation is well-optimised and introduces only a modest
overhead.

\section{Related Work}
\label{sec:relatedwork}

\mypar{Training with multiple GPUs} Machine learning systems use data
parallelism, model parallelism, or a mix or both (see
DistBelief~\cite{Dean2012} and, more recently, FlexFlow~\cite{jia18layerwise})
to decrease training time.
TensorFlow~\cite{Abadi2016osdi}, PyTorch~\cite{pytorch},
MXNet~\cite{DBLP:journals/corr/ChenLLLWWXXZZ15}, CNTK~\cite{DBLP:conf/kdd/SeideA16} and
Caffe2~\cite{GoyalDGNWKTJH17} exploit data parallelism by default and use S-SGD
as their de-facto training algorithm. S-SGD, however, couples the batch size
and the number of GPUs. To compensate for the loss of statistical efficiency
incurred by large batch sizes, users tune other hyper-parameters using
techniques such as auto-tuning~\cite{you2017scaling,tencent2018},
scaling~\cite{DBLP:journals/corr/Krizhevsky14} and warming
up~\cite{GoyalDGNWKTJH17,DBLP:journals/corr/Krizhevsky14} the learning rate,
auto-tuning the batch size~\cite{Smith2017},
auto-tuning the momentum~\cite{zhang2017yellowfin} and others. The
effectiveness of these tuning techniques is problem-specific~\cite{Dean2018},
and users invest substantial time to find a scalable
set-up~\cite{tencent2018}. \sys explores a different direction by decoupling
the batch size and the number of GPUs. It provides a design for a task engine
that can fully utilise a multi-GPU server even when the batch size is small.

\mypar{Increasing GPU utilisation} There are proposals to improve the hardware
utilisation of machine learning systems using cooperative scheduling.
ModelBatch~\cite{Narayanan2018} and NVIDIA's Multi-Process Service train
multiple deep learning models on a GPU in a cooperative manner, but the
problem of model synchronisation remains unresolved. Litz~\cite{Qiao2018}
explores the scheduling of training and synchronisation tasks on CPUs, but its
lack of GPU support makes it ineffective for deep learning.
Ray~\cite{Moritz2018} trains deep learning models using cooperative GPU tasks
but only shares GPUs using time-sharing. In contrast, \sys provides efficient
concurrent execution of learning and synchronisation tasks on GPUs, which is
the key to achieve high hardware efficiency when training deep learning models
with small batch sizes.

\mypar{Asynchronous training} Prior work also attempts to improve hardware
utilisation using asynchronous~SGD~\cite{Chaturapruek2015}, often at the
expense of statistical efficiency~\cite{Chen2016}. Hogwild!~\cite{Recht2011}
and Dogwild!~\cite{Noel2014} do not specify data dependencies between learning
and synchronisation tasks:
all workers access a central model concurrently, leading to higher training
throughput. To compensate for the loss in statistical efficiency,
DimmWitted~\cite{Zhang2014b} coordinates parallel replicas in a CPU-oriented
NUMA architecture. Each NUMA node has its own model replica shared by its
cores. Within a node, cores update a shared replica asynchronously. In contrast
to these efforts, \sys follows a synchronous training approach and therefore
does not compromise statistical efficiency with stale updates.

\myparr{Model averaging} was originally proposed as an asynchronous method to
distribute training. Polyak-Ruppert's \emph{averaged SGD}~\cite{Polyak1992,
  Polyak1990, Ruppert1988} first demonstrated that an average model can
asymptotically converge faster to a solution than the individual model replicas
used to compute it. In practice, it is difficult to find this asymptotic
region~\cite{Xu2011}, especially with models that have complex loss
spaces~\cite{Choromanska2018}. To improve the statistical efficiency of model
averaging, recent studies~\cite{Boyd2011,Li2014b} propose to use the average
model to correct the trajectory of model replicas, but the effectiveness of
this approach was shown only for non-deep-learning problems.

In deep learning, \emph{elastic averaging SGD}~(EA-SGD)~\cite{Zhang2015} uses
the average model to correct model replicas occasionally, keeping the
communication cost low.
\emph{Asynchronous decentralised SGD}~(AD-SGD)~\cite{lian18} further reduces
server communication traffic by requiring each replica to perform model
averaging with only one worker per iteration. Compared to these techniques,
\sync is a synchronous algorithm that shares and maintains a consistent view of
the central average model across all learners in each iteration. \sync further
improves the statistical efficiency of model averaging by adopting
momentum~(see~\Cref{sec:algorithm_design}) to correct the average model. When
there are changes to hyper-parameters during training, \sync also restarts the
averaging process as an effective way to preserve statistical efficiency.

\mypar{Distributed training} When scaling the training of deep learning models
in distributed clusters, a parameter server~(PS)~\cite{Li2014} design is the
de-facto approach. In contrast to \sys, which improves the training performance
with small batch sizes on a single multi-GPU server, PS-based systems address
the challenges of using a cluster for distributed learning, including the
handling of elastic and heterogeneous resources~\cite{Huang2018, jiang2017},
the mitigation of stragglers~\cite{Dean2012, cui2014, Harlap2016}, the
acceleration of synchronisation using hybrid hardware~\cite{Cui2016eurosys},
and the avoidance of resource fragmentation using collective
communication~\cite{Watcharapichat2016, sergeev2018horovod,
  tencent2018}. Similar to prior model averaging systems~\cite{Zhang2015}, \sys
could adopt a PS design to manage its average model in a distributed
deployment. We view the distribution of \sys as future work.

\section{Conclusions}
\label{sec:conclusions}

\sys improves hardware efficiency when training with the preferred batch size,
however small, with a low loss of statistical efficiency.
It trains multiple model replicas on the same \gpu, tuning their number
automatically as to maximise training throughput.
Despite training many more model replicas 
compared to existing approaches,
\sys avoids reduced statistical efficiency using \sync. The latter is 
a new training
algorithm in which replicas \emph{independently} explore the solution space
with gradient descent, but adjust their search \emph{synchronously} based on
the trajectory of a globally-consistent central average model.
Our experimental results with a set of deep learning models show that \sys
shortens the time-to-accuracy during training by up to
4$\times$ compared to TensorFlow.

\end{document}